%% file: ms.tex
\documentclass[12pt]{elsarticle}

\usepackage{amssymb}

\journal{Journal of Computational Physics: X}

\input{packages}

\colorlet{Reviewer1}{black}
\colorlet{Reviewer2}{black}

\colorlet{Reviewer12}{black}
\colorlet{Reviewer22}{black}

\colorlet{Reviewer13}{black}
\colorlet{Reviewer23}{black}

\colorlet{Me22}{orange}

\makeatletter
\def\ps@pprintTitle{%
 \let\@oddhead\@empty
 \let\@evenhead\@empty
 \def\@oddfoot{}%
 \let\@evenfoot\@oddfoot}
\makeatother

\begin{document}

\frenchspacing

\begin{frontmatter}


\author{Tomislav Mari\'{c}\corref{cor1}}
\address{Mathematical Modeling and Analysis, Technische Universit\"{a}t Darmstadt}
\cortext[cor1]{maric@mma.tu-darmstadt.de}

\title{Iterative Volume-of-Fluid interface positioning in general polyhedrons with Consecutive Cubic Spline interpolation}


\author{}

\address{}

\input{sections/abstract.tex}




\end{frontmatter}

\input{sections/introduction.tex}

\input{sections/ccs_positioning.tex}
\input{sections/results.tex}
\input{sections/conclusions.tex}



\bibliographystyle{elsarticle-num-names} 
\bibliography{literature}


\end{document}

%% file: packages.tex
\usepackage{hyperref}
\usepackage{url}
\usepackage{floatrow}
\usepackage{algorithm}
\usepackage{graphicx}
\usepackage{algpseudocode}
\usepackage{amsmath}
\usepackage[size=footnotesize]{subcaption}
\usepackage[textsize=tiny]{todonotes}
\graphicspath{{figures/}}
\usepackage{cleveref}
\usepackage{soul}
\usepackage{geometry}
\usepackage{algorithmicx}
\usepackage{booktabs}
\usepackage{longtable}
\usepackage{stmaryrd}
\newcommand{\irchi}[2]{\raisebox{\depth}{$#1\chi$}} 
\DeclareRobustCommand{\rchi}{{\mathpalette\irchi\relax}}
\newcommand{\x}{\ensuremath{\mathbf{x}}}

\newcommand{\xMax}{\ensuremath{\mathbf{x}_{max}}}
\newcommand{\xMin}{\ensuremath{\mathbf{x}_{min}}}
\newcommand{\p}{\ensuremath{\mathbf{p}}}
\newcommand{\sMax}{s_{max}}
\newcommand{\sMin}{s_{min}}
\newcommand{\n}{\ensuremath{\mathbf{n}}}

\newcommand{\VolFrac}{\ensuremath{\alpha}}
\newcommand{\VolFracStar}{\ensuremath{\VolFrac^{*}}}
\newcommand{\VolFracTilde}{\ensuremath{\tilde{\VolFrac}}}
\newcommand{\VolFracSet}{\ensuremath{\{\VolFrac_c\}_{c \in C}}}
\newcommand{\NormalSet}{\ensuremath{\{\n_c\}_{c \in C}}}

\newcommand{\ReconTol}{\ensuremath{\epsilon_R}}

%% file: sections/abstract.tex
\begin{abstract}

\noindent\textbf{This is the preprint version of the published manuscript \url{https://doi.org/10.1016/j.jcpx.2021.100093}: please cite the published manuscript when refering to the contents of this document.}

A straightforward and computationally efficient Consecutive Cubic Spline (CCS) iterative algorithm is proposed \textcolor{Reviewer12}{for positioning} the planar interface of the unstructured geometrical Volume-of-Fluid method in arbitrarily-shaped cells. \textcolor{Reviewer12}{The CCS algorithm is a two-point root-finding algorithm \citep[chap. 2]{Petkovic2012}, designed for the VOF interface positioning problem, where the volume fraction function has diminishing derivatives at the ends of the search interval. As a two-point iterative algorithm, CCS re-uses function values and derivatives from previous iterations and does not rely on interval bracketing. The CCS algorithm requires only two iterations on average to position the interface with a tolerance of $10^{-12}$, even with numerically very challenging volume fraction values, e.g., near $10^{-9}$ or $1-10^{-9}$.}

\textcolor{Reviewer12}{The proposed CCS algorithm is very straightforward to implement because its input is already calculated by every geometrical VOF method. It builds upon and significantly improves the predictive Newton method \citep{Chen2019} and is independent of the cell's geometrical model and related intersection algorithm. Geometrical parameterizations of truncated volumes used by other contemporary methods \citep{Diot2014, Diot2016, Lopez2018, Lopez2019} are completely avoided. The computational efficiency is comparable in terms of the number of iterations to the fastest methods reported so far. References are provided in the results section to the open-source implementation of the CCS algorithm and the performance measurement data.}
\end{abstract}

%% file: sections/introduction.tex
\section{Introduction}
\label{sec:intro}

\begin{figure}[!htb] 
    \footnotesize
    \centering
    \begin{subfigure}[t]{0.49\textwidth}
        \centering
        \def\svgwidth{\columnwidth}
        { 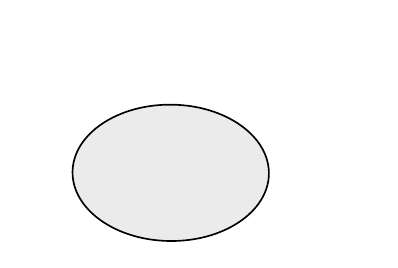 }
        \caption{Multiphase domain.}
        \label{fig:domain}
    \end{subfigure}
    \begin{subfigure}[t]{0.49\textwidth}
        \centering
        \def\svgwidth{\columnwidth}
        { 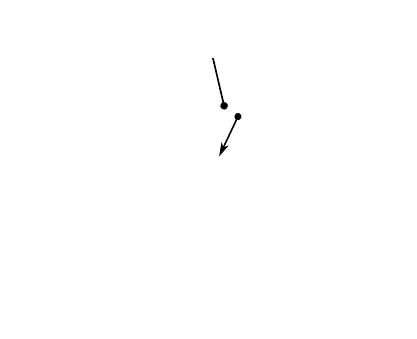 }
        \caption{Discretized multiphase domain.}
        \label{fig:domaindiscr}
    \end{subfigure}
    \caption{Multiphase domain and its discretization with the Volume-of-Fluid method.}
\end{figure} 

An essential task of any multiphase simulation method is the identification and tracking of immiscible fluid phases, schematically shown in \cref{fig:domain}. Each phase with distinctive physical properties can be associated with a subset $\Omega^{\pm}(t)$ of the domain $\Omega = \Omega^+(t) \cup \Omega^-(t)$, separated by the fluid interface $\Sigma(t)$ with outward-oriented normal vectors $\n_\Sigma$. An indicator function is introduced to distinguish $\Omega^\pm(t)$, as
\begin{equation}
    \rchi(t, \x) := 
        \begin{cases}
            1 \text{ if } \x \in \Omega^+(t) \\
            0 \text{ if } \x \in \Omega^-(t). 
        \end{cases}
    \label{eq:indicator}
\end{equation}
%
\textcolor{Reviewer13}{In numerical simulations of two-phase flows, approximations such as those shown in \cref{fig:domaindiscr} are used, specifically by the geometrical Volume-of-Fluid method, to approximately calculate $\Omega^\pm(t)$}. The domain $\Omega$ is discretized as a union of non-overlapping polyhedrons $V_c$, such that $\Omega=\cup_c V_c, c \in C$. For each polyhedron $V_c$, a \emph{volume fraction} is defined as
\begin{equation}
    \VolFrac_{c}(t) := \int_{V_c} \rchi(t, \x) \, \textcolor{Reviewer12}{dV},
    \label{eq:volfracdef}
\end{equation}
or, in other words, a \emph{fill level} of the polyhedron $V_c$ with $0 \le \VolFrac_c \le 1$. \textcolor{Reviewer2}{The Piecewise Linear Interface Calculation (PLIC) \citep{DeBar1974,Youngs1982,Rider1998} is still prevalently used by the geometrical VOF method to approximate the indicator function $\rchi(t, \x)$ in each $V_c$ using} 
\begin{equation}
    \rchi(t, \x) \approx H_{c} := H_{c}(\p_{c}(t), \n_{c}(t)) = 
    \begin{cases} 
        1 \text{ if } (\x - \p_{c}(t)) \cdot \n_{c}(t) \ge 0, \\
        0 \text{ otherwise.}
    \end{cases}
    \label{eq:hspaceapp}
\end{equation}
\textcolor{Reviewer2}{where $H_c(\p_c(t),\n_c(t))$ is the \emph{positive} halfspace of the PLIC plane given by the plane position $\p_c$ and plane normal $\n_c$ at time $t$. } The interface reconstruction algorithm computes the halfspace $H_{c}(\p_{c}(t), \n_{c}(t))$, by approximating the interface unit normal vector $\n_{c}$ and computing the point $\p_{c}$, such that the intersection between the positive halfspace and the polyhedron $V_c$ satisfies 
\begin{equation}
    \VolFrac_{c}(t) := \int_{V_c} \rchi(t, \x) \, \textcolor{Reviewer12}{dV} = \dfrac{|V_c \cap H_{c}(\p_{c}(t),\n_{c}(t))|}{|V_c|}. 
    \label{eq:volfrac}
\end{equation}
\textcolor{Reviewer2}{The computation of $\p_c$ in each $V_c$ is the \emph{interface positioning} part of the reconstruction algorithm, given $\NormalSet$, $\VolFracSet$, under the condition $0 < \VolFrac_c < 1$, and a polyhedral domain discretization $\Omega = \cup_c V_c, c \in C$. Since the goal of the interface positioning is to find $\p_c(t)$ at a fixed $t$, $t$ can be disregarded from now on.} 

\begin{figure}[!htb]
    \centering
    \begin{subfigure}[b]{0.49\textwidth}
        \centering
        \def\svgwidth{0.75\textwidth}
        {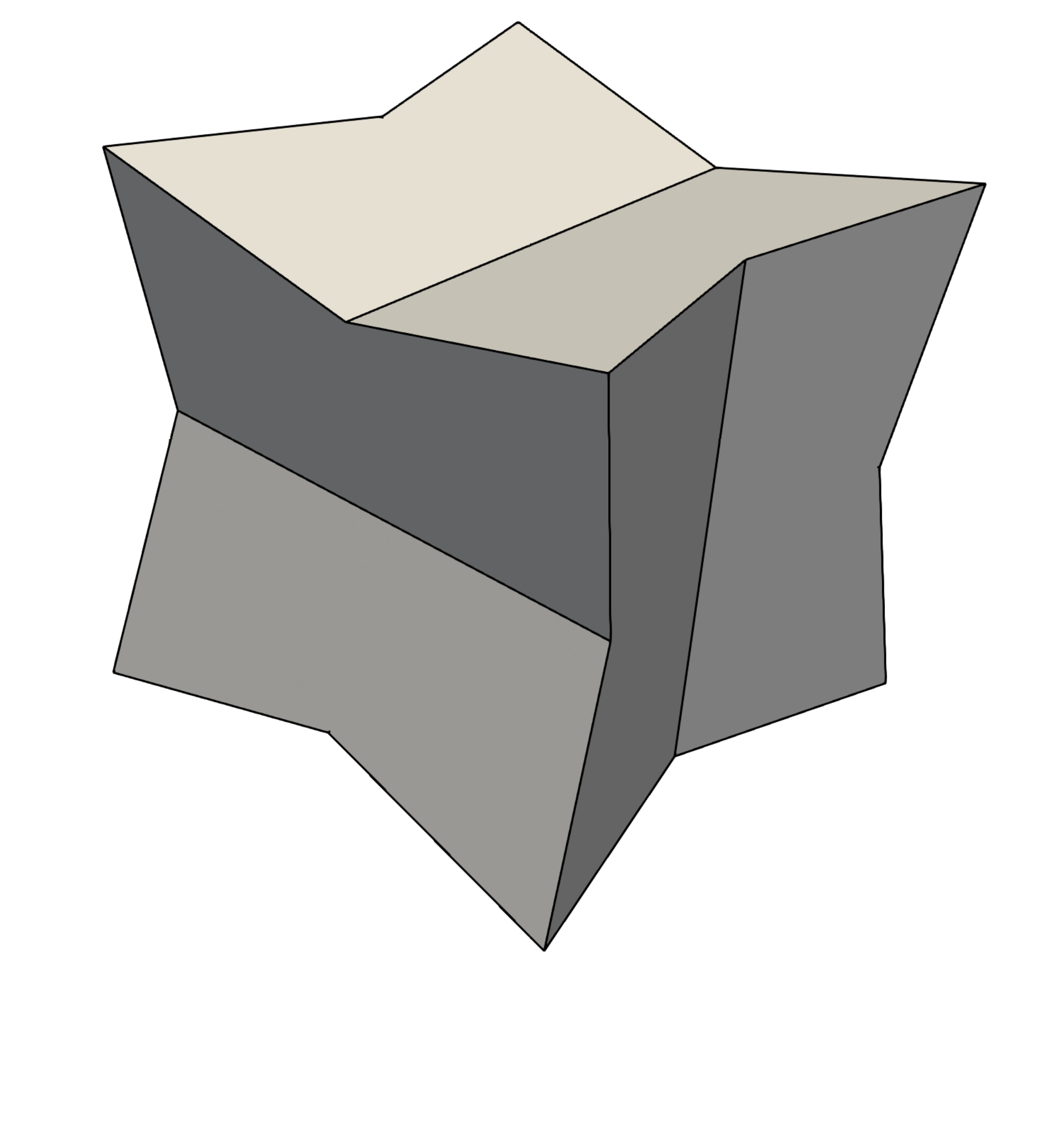 }
        \caption{Endo-dodecahedron.}
        \label{fig:pos-endo}
    \end{subfigure}
    \begin{subfigure}[b]{0.49\textwidth}
        \centering
        \def\svgwidth{\columnwidth}
        {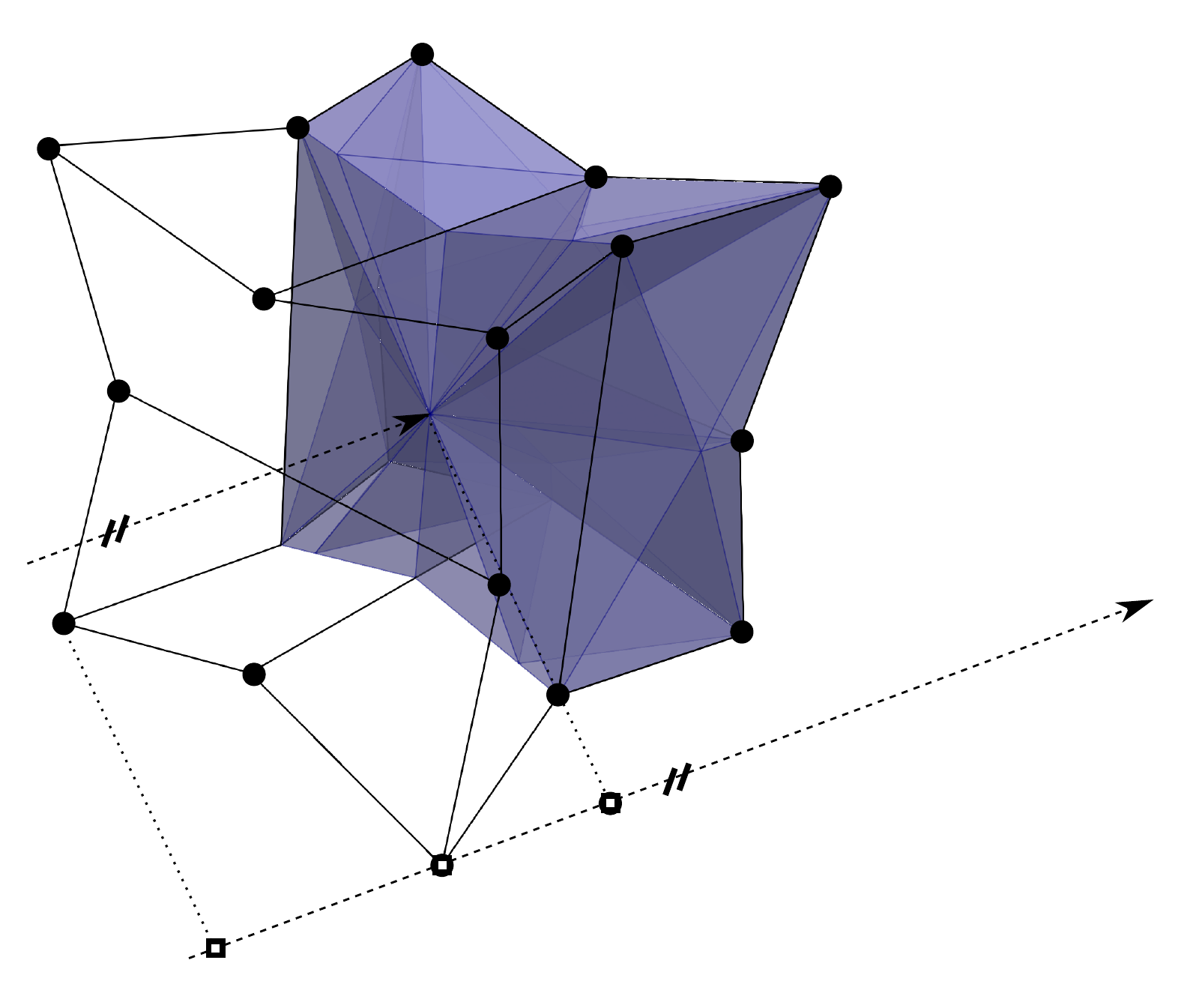}
        \caption{Intersected endo-dodecahedron.}
        \label{fig:pos-param}
    \end{subfigure}
    \caption{Interface positioning in an endo-dodecahedron.}
    \label{fig:pos}
\end{figure}

\textcolor{Reviewer2}{An explicit (closed-form) solution of \cref{eq:volfrac} for $\p_c$ does not exist for an arbitrary volume $V_c$. For example, the positioning of the PLIC plane in an endo-dodecahedron\footnote{A model for a non-convex dodecahedron with planar faces.} from \cref{fig:pos-endo} is shown in \cref{fig:pos-param} with $\VolFrac_c = 0.5$ and the normal vector $\n_c$ that is collinear with the $Z$-axis. The resulting intersection between the positive halfspace of the PLIC plane $H_c(\n_c, \p_c)$ and the endo-dodecahedron is the shaded volume in \cref{fig:pos-param}.}\textcolor{Reviewer2}{To simplify the solution process for $\p_c$ in \cref{eq:volfrac}, the position of $\p_c$ on the $\n_c$ axis can be parameterized with a scalar $s$, as shown in \cref{fig:pos-param}.}\textcolor{Reviewer1}{Minimal and maximal point $\x_{min,max}$ of the volume $V_c$, projected onto $\n_c$, as shown in \cref{fig:pos-param}, are defined as
\begin{equation}
    \begin{aligned}
        \x_{min} &= \x_{ref} + \min_p ((\mathbf{x}_p - \mathbf{x}_{ref}) \cdot \mathbf{n}_c) \cdot \n_c, \quad p \in P,\\ 
        \x_{max} &= \x_{ref} + \max_p ((\mathbf{x}_p - \mathbf{x}_{ref}) \cdot \mathbf{n}_c) \cdot \n_c, \quad p \in P, 
    \end{aligned}
    \label{eq:xmin} 
\end{equation}}\noindent\textcolor{Reviewer1}{where $\x_{ref}$ is a reference point for the $s$-axis. Any point can be chosen as $\x_{ref}$ and usually the origin of the coordinate system is used. However, using the origin as a reference point increases the error of floating-point calculations used in \cref{eq:xmin}, and using a point that belongs to $V_c$ (e.g. the first point, or the centroid of $V_c$) reduces these errors \cite[section 2]{Shewchuk2013}. The set $P$ in \cref{eq:xmin} is the set all points $x_p$ of the volume $V_c$.} Using \cref{eq:xmin}, the halfspace position $\p_c$ is parameterized with 
\begin{align}
    H_c(s) := H_c(\mathbf{p}_c(s), \mathbf{n}_c) \\
    \mathbf{p}_c(s) = \mathbf{x}_{min} + s \mathbf{n}_c,
    \label{eq:psparam}
\end{align}
where \textcolor{Reviewer1}{$s\in[\sMin, \sMax]$. Note that the parametrization determines the values of the parameter $s$: for example, with the $\xMin,\xMax$, parametrization by \cref{eq:psparam}, $s \in [0,s_{max}]$.} This, in turn, leads to the parametrization of the intersection volume in \cref{eq:volfrac} as 
\begin{equation}
    V(s) = V_c \cap H_c(\p_c(s), \n_c).
    \label{eq:vsparam}
\end{equation}
The parameterization of the intersection volume reformulates \cref{eq:volfrac} as
\begin{equation}
    \VolFracTilde_c(s) := \dfrac{|V(s)|}{|V_c|} - \VolFrac_c = \VolFrac_c(s) - \VolFrac_c = 0,
    \label{eq:volfracroot}
\end{equation}
where $\VolFrac_c$ is the given volume fraction for $V_c$. The root $s^*$ of $\VolFracTilde(s)$ is sought to compute $p_c(s^*)$, the position of the PLIC plane $(\mathbf{p}_c(s^*), \n_c)$ in $V_c$ such that \cref{eq:volfrac} is satisfied, as shown in \cref{fig:pos-param}.   

\begin{figure}[htb]
    \centering
    \footnotesize
    \begin{subfigure}[b]{0.3\textwidth}
        \centering
        \includegraphics[width=\columnwidth]{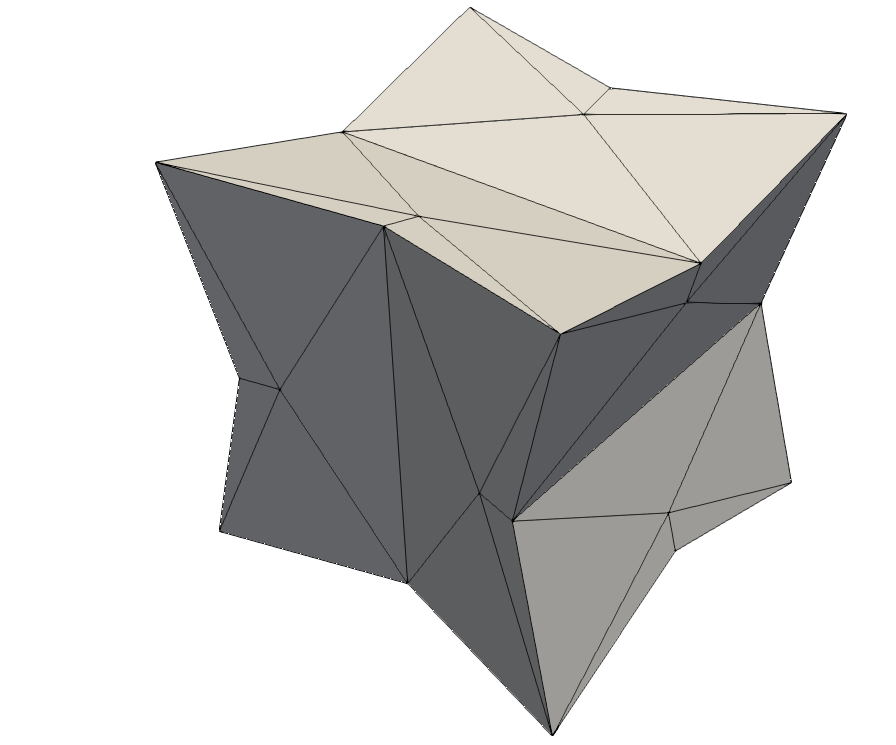}
        \caption{Endo-dodecahedron.}
        \label{fig:endo}
    \end{subfigure}
    \begin{subfigure}[b]{0.69\textwidth}
        \centering
        \includegraphics{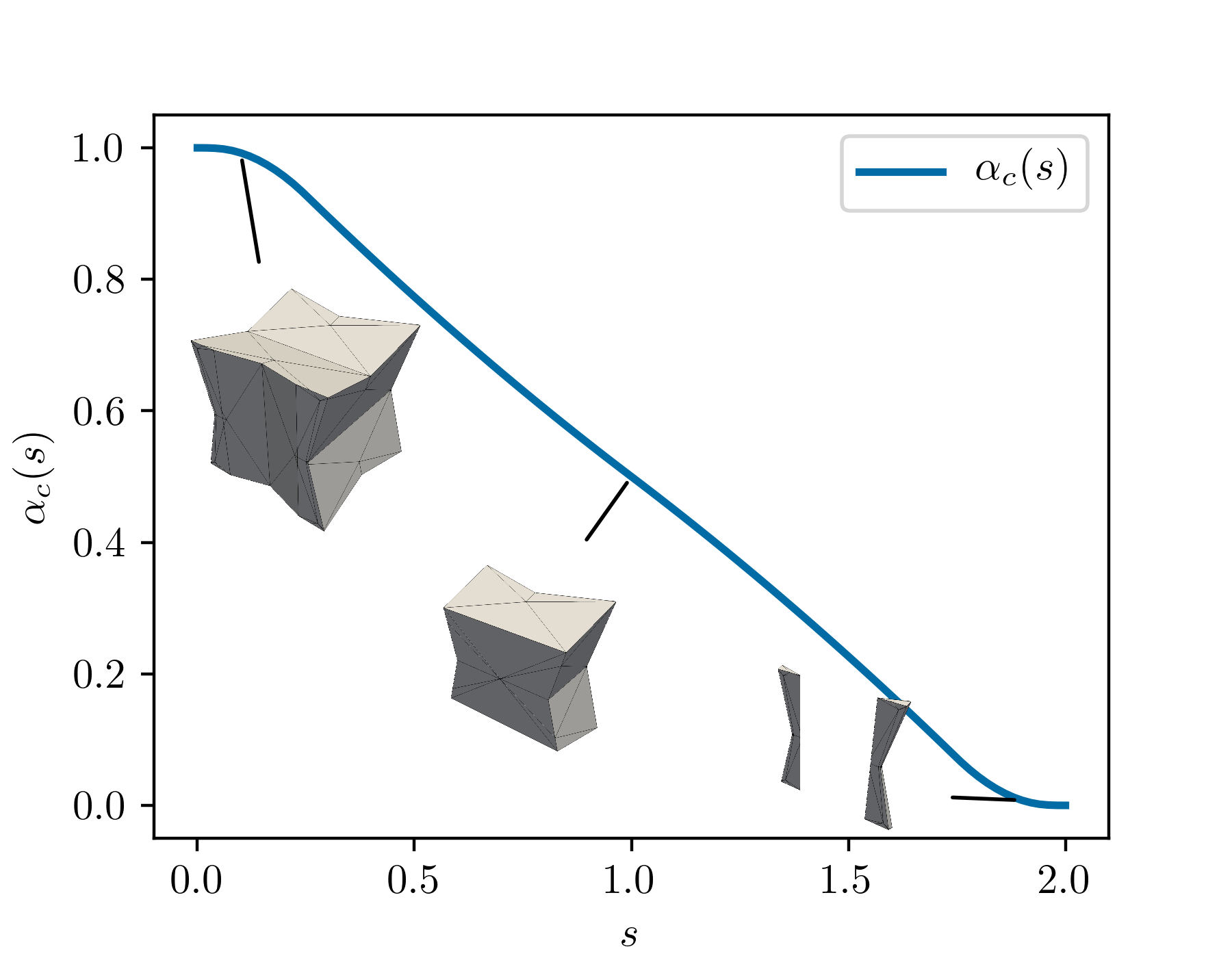}
        \caption{Endo-dodecahedron volume fraction $\VolFrac_c(s)$.}
        \label{fig:endofill}
    \end{subfigure}
    \caption{Geometry and the volume fraction $\VolFrac_c(s)$ of the endo-dodecahedron.}
    \label{fig:endofull}
\end{figure}

\textcolor{Reviewer2}{The function $\VolFrac(s)$ is visualized in \cref{fig:endofull} for the endo-dodecahedron shown in \cref{fig:pos-endo}}, using $\n_c = (0,0,1)$ and $\VolFracStar \in [0,\VolFracStar_1, \VolFracStar_2, \dots, \VolFracStar_i, 1], \VolFracStar_i = \frac{i}{N}$, $N=100$. If $V_c$ is non-convex, \textcolor{Reviewer2}{interface positioning sometimes separates $V_c$ into disjoint volumes, similar to the result shown in \cref{fig:endofill} for $s\approx 2$.} Geometrical operations used for the intersection must therefore support disjoint sets. \textcolor{Reviewer12}{Non-convex cells with non-planar faces are decomposed into tetrahedrons using the centroid of the cell and centroids of the non-planar faces. The magnitude of the truncated volume that lies inside the halfspace $H_c(\p_c(s), \n_c)$ is, therefore, equal to the sum of volume magnitudes of the truncated tetrahedrons. The tetrahedral decomposition is the most straightforward approach  to the volume truncation of non-convex cells with non-planar faces \citep{Ahn2007}, but it is more computationally expensive, compared to sophisticated approaches such as \citep{Lopez2019}. The choice of the truncation is arbitrary for the proposed CCS method: an alternative volume truncation reduces the total CPU time, however, it has no influence on the total number of iterations, whose very straightforward reduction the main contribution of the proposed method.}

\textcolor{Reviewer2}{Following boundary conditions are given for $\VolFracTilde(s)$:
\begin{align}
    \VolFracTilde(\sMin) & = 1 - \VolFrac_c,\\
    \VolFracTilde'(\sMin) & = 0, \\
    \VolFracTilde(\sMax) & = -\VolFrac_c, \\
    \VolFracTilde'(\sMax) & = 0. 
    \label{eq:volfraccond}
\end{align}}Boundary conditions given by \cref{eq:volfraccond} complicate the \textcolor{Reviewer2}{root finding in \cref{eq:volfracroot}}, because the derivatives of $\VolFracTilde(s)$ are diminishing at $s=\sMin, \sMax$. \textcolor{Reviewer12}{Furthermore, although highly improbable, it is possible that the normal vector $\n_c$ becomes collinear with the normal of a \emph{planar face} of the cell. In this case either $\VolFracTilde'(s_{min})\ne0$ or $\VolFracTilde'(s_{max})\ne0$. This special case of the boundary condition is handled by the proposed CCS method and addressed in detail below.} 

Positioning algorithms can be categorized as \emph{iterative} or \emph{bracketing} algorithms. Iterative algorithms rely solely on root-finding to solve \cref{eq:volfracroot} up to a prescribed tolerance. Bracketing algorithms intersect the volume $V_c$ incrementally until a \emph{bracketed volume} is found that contains the interface. \textcolor{Reviewer12}{In the bracketed volume, modern positioning algorithms \citep{Diot2016,Lopez2018,Lopez2019} \emph{exactly} position the interface using an exact function for the volume fraction within the bracketed interval.}

\citet{Kothe1996,Rider1998} have proposed an iterative algorithm that relied on Brent's root finding method, combining inverse quadratic interpolation with bisection to avoid divergence near interval boundaries. Bracketing intersects the volume $V_c$ with the halfspace given by $H(\p_p, \n_c), p \in P$, where $\{\p_p\}_{p \in P}$ are cell corner points. This brackets a volume $V_c$ as
\begin{equation}
    B_{c,i,j} = H(\p_i, \n_c) \cap V_c \setminus H(\p_j, \n_c) \cap V_c, \quad s_i \le s^* \le s_j, \p_{i,j} \in P,
    \label{eq:bracketvol}
\end{equation}
\textcolor{Reviewer2}{resulting in the bracketed volume $B_{c,i,j}$ that contains the root $s^*$ of $\VolFracTilde$ given by \cref{eq:volfracroot}.} Within the bracketed volume, Brent's algorithm requires additional iterations to find the root of \cref{eq:volfracroot} up to a prescribed tolerance. This procedure was widely adopted for interface positioning \citep{Shahbazi2003,Pilliod2004,Lopez2004,Liovic2006}. \citet{Scardovelli2000} have proposed an analytical interface positioning method for rectangular meshes that (used in \citep{Aulisa2003}) and \citet{Yang2006} have proposed an analytical positioning method for tetrahedral and triangular meshes. 

\citet{Lopez2008} have extended the work of \citet{Scardovelli2000} and \citet{Yang2006}. They have used and indexed face-set\footnote{An indexed face-set models a volume using a global set of unique points and its polygonal boundary as a set of polygons, such that each boundary polygon is modeled as a sequence of indexes of unique points.} as the boundary representation of a polyhedron \textcolor{Reviewer13}{(see \citep[Chapter 28]{Ghali2008} for details)} and an analytical expression for a volume of a convex polyhedron \citep{Schneider2002} to position the interface exactly within the bracketed interval. \textcolor{Reviewer1}{\citet{Lopez2008} introduce the Central Sequential Bracketing (CSB) procedure that utilizes sorted signed distances associated with the corner points $i_p$ of the volume $V_c$, calculated using the interface normal $\n_c$. The goal of the algorithm is to locate indices $k_{min}, k_{max}$ in this list, such that $k_{max} - k_{min} = 1$, $|V_{T,k_{max}}| \ge \VolFrac_c |V_c|$ and $|V_{T,k_{min}}| \le \VolFrac_c |V_c|$, where $V_{T,k}$ is the truncated volume given by the plane normal $\n_c$ and the vertex $i_p$ associated with the $k$-th signed distance in the list. The algorithm starts with the central index $k_c = INT[(I_p + 1)/2]$ in the signed distance list, and computes the truncated volume $V_T$ passing through this point. If $|V_T| > \VolFrac_c |V_c|$, $k_{max} = k_c$, otherwise $k_{min} = k_c$. The next iteration continues with the reduced list of indices between $[k_{min}, k_{max}]$. The authors state that the algorithm complexity in terms of the CPU time is $O(\frac{1}{4}I_p + 1)$ and $O(log_2 I_p)$ if so-called Binary Bracketing (BB) is used to set $k_c = (k_{min} + k_{max}) /2$, where $I_p$ is the total number of points of the volume $V_c$. They utilize BB for cells with $I_p > 8$ because then the logarithmic complexity outperforms the linear complexity of the CSB bracketing algorithm. \citet{Lopez2018} have published the source code that implements their methods from \citep{Lopez2008}.}

\textcolor{Reviewer1}{\citet{Ahn2008} have proposed a stabilized bisection-secant method as an iterative positioning algorithm, used by the author in \citep{Maric2013,Maric2018}. Their algorithm benefits from the super-linear (golden ratio) convergence of the secant method outside of the regions with diminishing derivatives, and relies on the bisection method to ensure convergence otherwise. They rely on the tetrahedral decomposition of the volume for the calculation of the truncated volume, however the root finding algorithm is independent of that choice.}

\textcolor{Reviewer1}{\citet{Diot2014,Diot2016} have proposed an analytical positioning method that does not rely on the analytical expression for the volume of the convex polygon (polyhedron). Instead, the polygon (polyhedron) is decomposed into sub-volumes whose magnitudes are exactly calculated using mixed vector and scalar products. The proposed method is very accurate, however it involves relatively complex subdivisions of the $V_c$ volume.}

\textcolor{Reviewer1}{\citet{Lopez2016} present a detailed review of interface positioning algorithms and propose an enhancement of their CSB and BB bracketing algorithms from \citep{Lopez2008} on convex polyhedral cells. Their new Interpolation Bracketing (IB) algorithm uses signed distances to linearly interpolate $H(\p_c(s), \n_c)$, in effect interpolating linearly the volume fraction function shown in \cref{fig:endofill}, \textcolor{Reviewer12}{which reduces the number of iterations of the bracketing algorithm}. The authors couple their IB algorithm with a new explicit function for interface positioning in the bracketed interval. The authors state that the explicit positioning function costs as much as $1.7$ times the volume truncation operation, which must be taken into account when its overall computational complexity is considered. The new explicit function is combined with the IB algorithm into the new Coupled Interpolation-Bracketed Analytical Volume Enforcement (CIBRAVE) method. }

\begin{figure}[!htb]
    \centering
    \footnotesize
    \def\svgwidth{0.7\textwidth}
    { 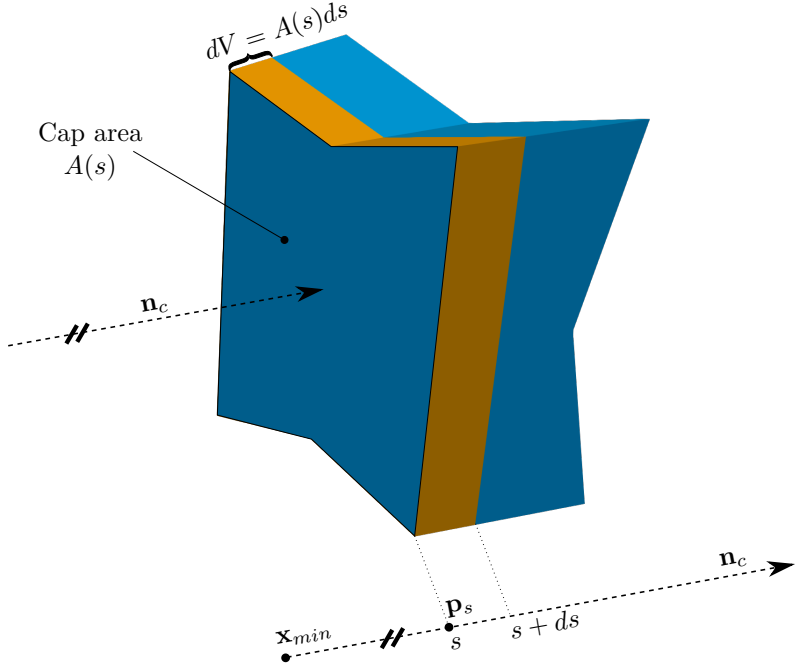 }
    \caption{\textcolor{Reviewer12}{Cap area} $A(s)$ used as the exact volume derivative $V'(s)$.}
    \label{fig:newtonpolyhedron}
\end{figure}

\textcolor{Reviewer2}{Recently, \citet{Chen2019} have noticed that the fundamental theorem of calculus can be used to increase the convergence of the iterative positioning algorithm. In \cref{fig:newtonpolyhedron}, the intersection of the plane $(\p(s), \n_c)$ and the volume $V_c$ is the polygon with the area $A(s)$ that lies on the plane $(\p(s), \n_c)$: the so-called \emph{cap polygon}. The area of this polygon (\textcolor{Reviewer12}{\emph{cap area}}) is computed geometrically, with machine-epsilon accuracy, at any $s$, and it can be used to integrate parametrized volume $V(s)$ as 
\begin{equation}
    V(s_2) = V(s_1) - \int_{s_1}^{s_2} A(s) ds = V(s_1) - \int_{s_1}^{s_2} V'(s) ds, \quad s_1 < s_2.
    \label{eq:areaint}
\end{equation}
It is obvious that $A(s) = V'(s)$ is the exact derivative of the volume $V(s)$ from \cref{eq:volfracroot}, whose root determines the position of the interface $\p(s^*)$. A direct consequence of this is 
\begin{equation}
    \VolFrac'(s) = \frac{|V'(s)|}{|V_c|} = \frac{|A(s)|}{|V_c|}.
\end{equation}
\citet{Chen2019} use the exact derivative $\VolFrac'(s)$ in the Newton's root finding method to solve \cref{eq:volfracroot}. Knowing the exact derivative ensures the quadratic convergence of the Newton's method, provided that the condition $\VolFrac'(s)\ne0$ is fulfilled. However, it does not solve issue of divergence of the Newton's method near interval boundaries $\sMin, \sMax$. If an iteration of the Newton's method computes a root outside of $(\sMin, \sMax)$, \citet[Algorithm 1]{Chen2019} apply the Hermite cubic spline interpolation at $s$, across $(s_{min}, s_{max})$. Unfortunately, interpolating across the whole search interval $(\sMin,\sMax)$ introduces large interpolation errors and substantially slows down convergence for volume fraction values $\VolFrac_c\approx 0, \VolFrac \approx 1$. To understand why, consider the case when $\VolFrac_c \approx 1$: the derivative $\VolFrac'(s)\approx 0$, Newton's method shoots out of the search interval $\sMin,\sMax$, and then Hermite interpolation over $\sMin,\sMax$ is performed. If $\VolFrac\approx 1$, the root lies near $\sMin$. However, the Hermite interpolation introduces the conditions on $\VolFracTilde(s)$ at $\sMax$, that lies on the other end of the search interval. Therefore, the accuracy is lost in this case, and vice versa for $\VolFrac_c \approx 0$. The method of \citet{Chen2019} is named Newton Cubic Spline (NCS) method to facilitate comparison with the proposed method.}

\textcolor{Reviewer2}{Similar to Brent's method, which is a multipoint method\footnote{Multipoint root finding methods re-use function values and derivative from previous iterations to increase the convergence order.}, the algorithm proposed here reuses function values $\VolFracTilde(s)$ and derivatives $\VolFracTilde'(s)$ from previous iterations to effectively double the \textcolor{Reviewer12}{order of accuracy of the interpolated volume fraction}. This is the basis of the proposed Consecutive Cubic Spline (CCS) interpolation. An additional stabilization for the ends of the interval $[\sMin,\sMax]$ is developed, that effectively handles diminishing derivatives at the ends of the search interval.}

\textcolor{Reviewer1}{\citet{Chen2019} increase the computational efficiency of the Moment-of-Fluid Method \citep{Dyadechko2005}, by expressing the centroid of the cap polygon $A(s)$ as a function of spherical interface orientation angles $\theta, \phi$. This is relevant for increasing the efficiency of interface positioning in the context of algorithms that improve the orientation of the VOF interface. The CCS algorithm proposed in this manuscript achieves significantly higher computational efficiency compared to NCS, especially for challenging values $\VolFrac_c \approx 1, \VolFrac_c \approx 0$, even without utilizing the information about the interface orientation. Of course, CCS can be coupled to MoF or any other algorithm that improves interface orientation in the same way as NCS. Since the convergence order of the CCS algorithm, demonstrated in the following section, is significantly higher than NCS without the use of interface orientation, adding this information would only increase the computational efficiency of the interface reconstruction algorithm. In future work, CCS will be coupled with the simplified Swartz reconstruction algorithm \citep{Maric2018}.}

%% file: figures/mathmodel-domain.pdf_tex
\begingroup%
  \makeatletter%
  \providecommand\color[2][]{%
    \errmessage{(Inkscape) Color is used for the text in Inkscape, but the package 'color.sty' is not loaded}%
    \renewcommand\color[2][]{}%
  }%
  \providecommand\transparent[1]{%
    \errmessage{(Inkscape) Transparency is used (non-zero) for the text in Inkscape, but the package 'transparent.sty' is not loaded}%
    \renewcommand\transparent[1]{}%
  }%
  \providecommand\rotatebox[2]{#2}%
  \newcommand*\fsize{\dimexpr\f@size pt\relax}%
  \newcommand*\lineheight[1]{\fontsize{\fsize}{#1\fsize}\selectfont}%
  \ifx\svgwidth\undefined%
    \setlength{\unitlength}{190.58421707bp}%
    \ifx\svgscale\undefined%
      \relax%
    \else%
      \setlength{\unitlength}{\unitlength * \real{\svgscale}}%
    \fi%
  \else%
    \setlength{\unitlength}{\svgwidth}%
  \fi%
  \global\let\svgwidth\undefined%
  \global\let\svgscale\undefined%
  \makeatother%
  \begin{picture}(1,0.67256949)%
    \lineheight{1}%
    \setlength\tabcolsep{0pt}%
    \put(0,0){\includegraphics[width=\unitlength,page=1]{mathmodel-domain.pdf}}%
    \put(0.05992198,0.48996575){\color[rgb]{0,0,0}\makebox(0,0)[lt]{\lineheight{0}\smash{\begin{tabular}[t]{l}$\Sigma(t)$\end{tabular}}}}%
    \put(0,0){\includegraphics[width=\unitlength,page=2]{mathmodel-domain.pdf}}%
    \put(0.64185011,0.52298277){\color[rgb]{0,0,0}\makebox(0,0)[lt]{\lineheight{0}\smash{\begin{tabular}[t]{l}$\n_\Sigma$\end{tabular}}}}%
    \put(0.39575763,0.22882148){\color[rgb]{0,0,0}\makebox(0,0)[lt]{\lineheight{0}\smash{\begin{tabular}[t]{l}$\Omega^+(t)$\end{tabular}}}}%
    \put(0.66008048,0.08321654){\color[rgb]{0,0,0}\makebox(0,0)[lt]{\lineheight{0}\smash{\begin{tabular}[t]{l}$\Omega^-(t)$\end{tabular}}}}%
    \put(0,0){\includegraphics[width=\unitlength,page=3]{mathmodel-domain.pdf}}%
    \put(0.46756215,0.62024891){\color[rgb]{0,0,0}\makebox(0,0)[lt]{\lineheight{0}\smash{\begin{tabular}[t]{l}$\Omega$\end{tabular}}}}%
    \put(0.29095859,0.4364679){\color[rgb]{0,0,0}\makebox(0,0)[lt]{\lineheight{0}\smash{\begin{tabular}[t]{l}$\rchi(t, \cdot) = 0$\end{tabular}}}}%
    \put(0.25975172,0.14867316){\color[rgb]{0,0,0}\makebox(0,0)[lt]{\lineheight{0}\smash{\begin{tabular}[t]{l}$\rchi(t, \cdot) = 1$\end{tabular}}}}%
    \put(-0.04820455,0.52363097){\color[rgb]{0,0,0}\makebox(0,0)[lt]{\begin{minipage}{1.16454114\unitlength}\raggedright \end{minipage}}}%
  \end{picture}%
\endgroup%

%% file: figures/mathmodel-domain-discrete.pdf_tex
\begingroup%
  \makeatletter%
  \providecommand\color[2][]{%
    \errmessage{(Inkscape) Color is used for the text in Inkscape, but the package 'color.sty' is not loaded}%
    \renewcommand\color[2][]{}%
  }%
  \providecommand\transparent[1]{%
    \errmessage{(Inkscape) Transparency is used (non-zero) for the text in Inkscape, but the package 'transparent.sty' is not loaded}%
    \renewcommand\transparent[1]{}%
  }%
  \providecommand\rotatebox[2]{#2}%
  \newcommand*\fsize{\dimexpr\f@size pt\relax}%
  \newcommand*\lineheight[1]{\fontsize{\fsize}{#1\fsize}\selectfont}%
  \ifx\svgwidth\undefined%
    \setlength{\unitlength}{199.98321533bp}%
    \ifx\svgscale\undefined%
      \relax%
    \else%
      \setlength{\unitlength}{\unitlength * \real{\svgscale}}%
    \fi%
  \else%
    \setlength{\unitlength}{\svgwidth}%
  \fi%
  \global\let\svgwidth\undefined%
  \global\let\svgscale\undefined%
  \makeatother%
  \begin{picture}(1,0.81178405)%
    \lineheight{1}%
    \setlength\tabcolsep{0pt}%
    \put(0.49662031,0.6847827){\color[rgb]{0,0,0}\makebox(0,0)[lt]{\lineheight{0}\smash{\begin{tabular}[t]{l}$H_c$\end{tabular}}}}%
    \put(0,0){\includegraphics[width=\unitlength,page=1]{mathmodel-domain-discrete.pdf}}%
    \put(0.50375719,0.3957533){\color[rgb]{0,0,0}\makebox(0,0)[lt]{\lineheight{0}\smash{\begin{tabular}[t]{l}$\n_c$\end{tabular}}}}%
    \put(0,0){\includegraphics[width=\unitlength,page=2]{mathmodel-domain-discrete.pdf}}%
    \put(0.45268557,0.7765077){\color[rgb]{0,0,0}\makebox(0,0)[lt]{\lineheight{0}\smash{\begin{tabular}[t]{l}$\Omega$\end{tabular}}}}%
    \put(0,0){\includegraphics[width=\unitlength,page=3]{mathmodel-domain-discrete.pdf}}%
    \put(0.59238768,0.53713266){\color[rgb]{0,0,0}\makebox(0,0)[lt]{\lineheight{0}\smash{\begin{tabular}[t]{l}$\p_c$\end{tabular}}}}%
    \put(0,0){\includegraphics[width=\unitlength,page=4]{mathmodel-domain-discrete.pdf}}%
    \put(0.12705784,0.60081273){\color[rgb]{0,0,0}\makebox(0,0)[lt]{\lineheight{0}\smash{\begin{tabular}[t]{l}$V_c$\end{tabular}}}}%
    \put(0.52129375,0.62133105){\color[rgb]{0,0,0}\makebox(0,0)[lt]{\begin{minipage}{0.00375031\unitlength}\raggedright \end{minipage}}}%
    \put(0.04916317,0.04564568){\color[rgb]{0,0,0}\makebox(0,0)[lt]{\lineheight{0}\smash{\begin{tabular}[t]{l}$\alpha_c = 1$\end{tabular}}}}%
    \put(0,0){\includegraphics[width=\unitlength,page=5]{mathmodel-domain-discrete.pdf}}%
    \put(0.33196567,0.04564568){\color[rgb]{0,0,0}\makebox(0,0)[lt]{\lineheight{0}\smash{\begin{tabular}[t]{l}$0 < \alpha_c < 1$\end{tabular}}}}%
    \put(0,0){\includegraphics[width=\unitlength,page=6]{mathmodel-domain-discrete.pdf}}%
    \put(0.70890978,0.04564568){\color[rgb]{0,0,0}\makebox(0,0)[lt]{\lineheight{0}\smash{\begin{tabular}[t]{l}$\alpha_c = 0$\end{tabular}}}}%
    \put(0,0){\includegraphics[width=\unitlength,page=7]{mathmodel-domain-discrete.pdf}}%
    \put(0.06477681,0.11097719){\color[rgb]{0,0,0}\makebox(0,0)[lt]{\begin{minipage}{0.06099309\unitlength}\raggedright \end{minipage}}}%
  \end{picture}%
\endgroup%

%% file: figures/positioning-parametrization-visualization-endo.pdf_tex
\begingroup%
  \makeatletter%
  \providecommand\color[2][]{%
    \errmessage{(Inkscape) Color is used for the text in Inkscape, but the package 'color.sty' is not loaded}%
    \renewcommand\color[2][]{}%
  }%
  \providecommand\transparent[1]{%
    \errmessage{(Inkscape) Transparency is used (non-zero) for the text in Inkscape, but the package 'transparent.sty' is not loaded}%
    \renewcommand\transparent[1]{}%
  }%
  \providecommand\rotatebox[2]{#2}%
  \newcommand*\fsize{\dimexpr\f@size pt\relax}%
  \newcommand*\lineheight[1]{\fontsize{\fsize}{#1\fsize}\selectfont}%
  \ifx\svgwidth\undefined%
    \setlength{\unitlength}{1125bp}%
    \ifx\svgscale\undefined%
      \relax%
    \else%
      \setlength{\unitlength}{\unitlength * \real{\svgscale}}%
    \fi%
  \else%
    \setlength{\unitlength}{\svgwidth}%
  \fi%
  \global\let\svgwidth\undefined%
  \global\let\svgscale\undefined%
  \makeatother%
  \begin{picture}(1,1.08266667)%
    \lineheight{1}%
    \setlength\tabcolsep{0pt}%
    \put(0,0){\includegraphics[width=\unitlength,page=1]{positioning-parametrization-visualization-endo.pdf}}%
  \end{picture}%
\endgroup%

%% file: figures/positioning-parametrization-visualization.pdf_tex
\begingroup%
  \makeatletter%
  \providecommand\color[2][]{%
    \errmessage{(Inkscape) Color is used for the text in Inkscape, but the package 'color.sty' is not loaded}%
    \renewcommand\color[2][]{}%
  }%
  \providecommand\transparent[1]{%
    \errmessage{(Inkscape) Transparency is used (non-zero) for the text in Inkscape, but the package 'transparent.sty' is not loaded}%
    \renewcommand\transparent[1]{}%
  }%
  \providecommand\rotatebox[2]{#2}%
  \newcommand*\fsize{\dimexpr\f@size pt\relax}%
  \newcommand*\lineheight[1]{\fontsize{\fsize}{#1\fsize}\selectfont}%
  \ifx\svgwidth\undefined%
    \setlength{\unitlength}{764.05169678bp}%
    \ifx\svgscale\undefined%
      \relax%
    \else%
      \setlength{\unitlength}{\unitlength * \real{\svgscale}}%
    \fi%
  \else%
    \setlength{\unitlength}{\svgwidth}%
  \fi%
  \global\let\svgwidth\undefined%
  \global\let\svgscale\undefined%
  \makeatother%
  \begin{picture}(1,0.84461029)%
    \lineheight{1}%
    \setlength\tabcolsep{0pt}%
    \put(-0.07839146,0.9032911){\color[rgb]{0,0,0}\makebox(0,0)[lt]{\begin{minipage}{1.22590895\unitlength}\raggedright \end{minipage}}}%
    \put(0.02958582,0.20555844){\color[rgb]{0,0,0}\makebox(0,0)[lt]{\begin{minipage}{0.19459864\unitlength}\raggedright \end{minipage}}}%
    \put(0,0){\includegraphics[width=\unitlength,page=1]{positioning-parametrization-visualization.pdf}}%
    \put(0.90923419,0.36338119){\color[rgb]{0,0,0}\makebox(0,0)[lt]{\lineheight{1.25}\smash{\begin{tabular}[t]{l}$\n_c$\end{tabular}}}}%
    \put(0.96528163,0.30697752){\color[rgb]{0,0,0}\makebox(0,0)[lt]{\lineheight{1.25}\smash{\begin{tabular}[t]{l}$s$\end{tabular}}}}%
    \put(0.38393242,0.07866834){\color[rgb]{0,0,0}\makebox(0,0)[lt]{\lineheight{1.25}\smash{\begin{tabular}[t]{l}$\x_q$\end{tabular}}}}%
    \put(0.0983202,0.43021277){\color[rgb]{0,0,0}\makebox(0,0)[lt]{\lineheight{1.25}\smash{\begin{tabular}[t]{l}$\n_c$\end{tabular}}}}%
    \put(0.53420677,0.11936584){\color[rgb]{0,0,0}\makebox(0,0)[lt]{\lineheight{1.25}\smash{\begin{tabular}[t]{l}$\p_c(s^*)$\end{tabular}}}}%
    \put(0.3557169,0.81329852){\color[rgb]{0,0,0}\makebox(0,0)[lt]{\lineheight{1.25}\smash{\begin{tabular}[t]{l}$\x_p$\end{tabular}}}}%
    \put(0.18646098,0.01352015){\color[rgb]{0,0,0}\makebox(0,0)[lt]{\lineheight{1.25}\smash{\begin{tabular}[t]{l}$\x_{min}$\end{tabular}}}}%
  \end{picture}%
\endgroup%

%% file: figures/newton-cubic-spline-cap.pdf_tex
\begingroup%
  \makeatletter%
  \providecommand\color[2][]{%
    \errmessage{(Inkscape) Color is used for the text in Inkscape, but the package 'color.sty' is not loaded}%
    \renewcommand\color[2][]{}%
  }%
  \providecommand\transparent[1]{%
    \errmessage{(Inkscape) Transparency is used (non-zero) for the text in Inkscape, but the package 'transparent.sty' is not loaded}%
    \renewcommand\transparent[1]{}%
  }%
  \providecommand\rotatebox[2]{#2}%
  \newcommand*\fsize{\dimexpr\f@size pt\relax}%
  \newcommand*\lineheight[1]{\fontsize{\fsize}{#1\fsize}\selectfont}%
  \ifx\svgwidth\undefined%
    \setlength{\unitlength}{684.13128662bp}%
    \ifx\svgscale\undefined%
      \relax%
    \else%
      \setlength{\unitlength}{\unitlength * \real{\svgscale}}%
    \fi%
  \else%
    \setlength{\unitlength}{\svgwidth}%
  \fi%
  \global\let\svgwidth\undefined%
  \global\let\svgscale\undefined%
  \makeatother%
  \begin{picture}(1,0.86843085)%
    \lineheight{1}%
    \setlength\tabcolsep{0pt}%
    \put(0,0){\includegraphics[width=\unitlength,page=1]{newton-cubic-spline-cap.pdf}}%
    \put(0.17352714,0.46629945){\color[rgb]{0,0,0}\rotatebox{10.220324}{\makebox(0,0)[lt]{\lineheight{1.25}\smash{\begin{tabular}[t]{l}$\n_c$\end{tabular}}}}}%
    \put(0,0){\includegraphics[width=\unitlength,page=2]{newton-cubic-spline-cap.pdf}}%
    \put(0.90077188,0.14626471){\color[rgb]{0,0,0}\rotatebox{10.804855}{\makebox(0,0)[lt]{\lineheight{1.25}\smash{\begin{tabular}[t]{l}$\n_c$\end{tabular}}}}}%
    \put(0.56202873,0.04147092){\color[rgb]{0,0,0}\rotatebox{10.5734155}{\makebox(0,0)[lt]{\lineheight{1.25}\smash{\begin{tabular}[t]{l}$s$\end{tabular}}}}}%
    \put(0.34375831,0.0483665){\color[rgb]{0,0,0}\rotatebox{10.544743}{\makebox(0,0)[lt]{\lineheight{1.25}\smash{\begin{tabular}[t]{l}$\x_{min}$\end{tabular}}}}}%
    \put(0,0){\includegraphics[width=\unitlength,page=3]{newton-cubic-spline-cap.pdf}}%
    \put(0.10888922,0.680963){\color[rgb]{0,0,0}\makebox(0,0)[t]{\lineheight{1.25}\smash{\begin{tabular}[t]{c}Cap area \\$A(s)$\end{tabular}}}}%
    \put(0,0){\includegraphics[width=\unitlength,page=4]{newton-cubic-spline-cap.pdf}}%
    \put(0.63800925,0.05531022){\color[rgb]{0,0,0}\rotatebox{9.94678845}{\makebox(0,0)[lt]{\lineheight{1.25}\smash{\begin{tabular}[t]{l}$s+ds$\end{tabular}}}}}%
    \put(0,0){\includegraphics[width=\unitlength,page=5]{newton-cubic-spline-cap.pdf}}%
    \put(0.25643358,0.78237745){\color[rgb]{0,0,0}\rotatebox{17.434154}{\makebox(0,0)[lt]{\lineheight{1.25}\smash{\begin{tabular}[t]{l}$dV =A(s)ds$\end{tabular}}}}}%
    \put(0,0){\includegraphics[width=\unitlength,page=6]{newton-cubic-spline-cap.pdf}}%
    \put(0.55698431,0.08780249){\color[rgb]{0,0,0}\rotatebox{10.544743}{\makebox(0,0)[lt]{\lineheight{1.25}\smash{\begin{tabular}[t]{l}$\p_{s}$\end{tabular}}}}}%
  \end{picture}%
\endgroup%

%% file: sections/ccs_positioning.tex
\section{Iterative positioning with Consecutive Cubic Spline interpolation}
\label{sec:pos}

\textcolor{Reviewer1}{\citet{Chen2019} use $V(s_n),V'(s_n)=A(s_n)$ in a Newton iteration 
\begin{equation}
    s_{n+1} = s_n - \dfrac{\VolFracTilde(s_n)}{\VolFracTilde'(s_n)},
    \label{eq:newton}
\end{equation}
known to be second-order convergent if $\VolFracTilde'(s_n)\ne 0$ within the search interval, which is not the case for interface positioning, where the function derivatives diminish on the boundaries of the search interval.} 

The proposed Consecutive Cubic Spline (CCS) algorithm \textcolor{Reviewer12}{interpolates the volume fraction $\VolFracTilde(s)$ with a Hermite polynomial between two consecutive iteration steps when they contain the root, by re-using volume fraction values and derivatives from previous iterations.}

\textcolor{Reviewer12}{Using function value and function derivative from two subsequent iterations makes the CCS a two-point root finding algorithm \citep[chap. 2]{Petkovic2012}. If the interpolation interval does contain the root, the cubic interpolation exactly recovers the volume fraction $\VolFracTilde(s)$, and its root is therefore the root of $\VolFracTilde(s)$. The volume fraction $\VolFracTilde(s)$ is} approximated with the polynomial of order $K < 4$, $P_k(s) = \sum_{k=0 \dots K} a_k s^k \approx \VolFracTilde(s)$, using the conditions
\begin{equation}
    \begin{aligned}
    P_k(s_a) & = \VolFracTilde(s_a), \\
    P_k'(s_a) & = \VolFracTilde'(s_a), \\
    P_k(s_b) & = \VolFracTilde(s_b), \\
    P_k'(s_b) & = \VolFracTilde'(s_b), 
    \end{aligned}
    \label{eq:cubicpoly}
\end{equation}
that determine $k$, if $s^* \in [s_a,s_b] \subset (\sMin,\sMax)$, where $s^*$ is the root of $\VolFracTilde(s)$ given by \cref{eq:volfracroot}. 

To understand why $P_k(s)$ recovers the volume fraction exactly, consider the following. \textcolor{Reviewer12}{The volume $V_c$ is bounded by planar polygons or, if the boundary of $V_c$ consists of non-planar faces, by sets of triangles resulting from the triangulation of non-planar faces. In both cases}, the \textcolor{Reviewer12}{cap area} $A(\x=\x_{min} + s \n_c)$ in \cref{fig:newtonpolyhedron} is at most \textcolor{Reviewer13}{a quadratic function of the $(x,y,z)$ three-dimensional coordinates}. Because the parametrization $\x = \x_{min} + s \n_c$ given by \cref{eq:psparam} is linear, parametrized $A(s)$ is then also at most quadratic in $s$, in three dimensions. By \cref{eq:areaint}, $\VolFracTilde(s)$ is then at most cubic in three dimensions. We also know that $\VolFracTilde(s)$ is piecewise-polynomial between the points $\x_p$ of the volume $V_p$: it is $C^1$ continuous at the boundaries of intervals bracketed by the points $\x_p$ of $V_c$, because its derivative, the \textcolor{Reviewer12}{cap (intersection) area} $A(s)$, is $C^0$ continuous in $s$, \textcolor{Reviewer12}{in cells that are used in unstructured meshes for the discretization of Partial Differential Equations}. Therefore, 
\begin{equation}
    \VolFracTilde(s) \in \mathbb{P}^k, \quad k = 0,1,2,3,
    \label{eq:cubicfrac}
\end{equation}
inside the interval bracketed by points $\x_p$ of the volume $V_c$, and it is $C^1$ continuous on the boundaries of bracketed intervals. \textcolor{Reviewer12}{The $\VolFracTilde(s)$, defined by \cref{eq:volfracroot}, is calculated geometrically using the truncated volume $V_c$. Similarly, $\VolFracTilde'(s)$ is geometrically calculated as $A(s)$, the \textcolor{Reviewer12}{cap area} in \cref{fig:newtonpolyhedron}. Therefore, the conditions given by \cref{eq:cubicpoly} are satisfied exactly up to machine epsilon. \textcolor{Reviewer12}{Consequentially, $P_k(s)$ exactly interpolates $\VolFracTilde(s)$ within the bracketed interval and an exact root of $P_k(s)$ is equivalent to the root of $\VolFracTilde(s)$ within the bracketing interval.} Note that the calculation of $\VolFracTilde'(s)$ \emph{comes at no additional computational cost}, because $A(s)$ is a byproduct of the volume truncation used to compute $V(s)$ for $\VolFracTilde(s)$.}

\begin{figure}[tb]
    \footnotesize
    \centering
    \begin{subfigure}[t]{0.4\textwidth}
      \centering
      \def\svgwidth{\columnwidth}
        { 
            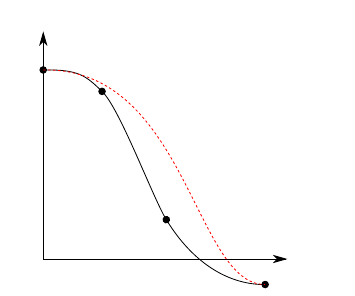 
        }
        \caption{\textcolor{Reviewer12}{CCS initial guess: interpolate $\VolFracTilde(s)$ over the initial search interval using boundary conditions given by \cref{eq:cubicpoly}.}}
        \label{fig:ccs1}
    \end{subfigure}
    \quad
    \begin{subfigure}[t]{0.4\textwidth}
      \centering
      \def\svgwidth{\columnwidth}
        { 
            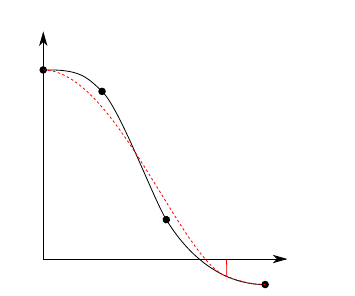 
        }
        \caption{\textcolor{Reviewer12}{CCS first iteration: a cubic spline interpolation using $[s_0, s_2, s_1]$ and \textcolor{Reviewer13}{$(\VolFracTilde(s_0), \VolFracTilde(s_1), \VolFracTilde(s_2), \VolFracTilde'(s_2))$}.}}
        \label{fig:ccs2}
    \end{subfigure}
    \begin{subfigure}[b]{0.6\textwidth}
      \centering
      \def\svgwidth{0.5\columnwidth}
        { 
            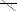 
        }
        \caption{\textcolor{Reviewer12}{CCS final iterations: $s_4$ is given by the Newton's method, because $s^* \notin [s_2, s_3]$. Since $s^* \in [s_4, s_3]$, the CCS polynomial is used over $[s_4, s_3]$ with \textcolor{Reviewer13}{$(\VolFracTilde(s_{4}),\VolFracTilde'(s_{4}),\VolFracTilde(s_{3}),\VolFracTilde'(s_{3}))$}.}}
        \label{fig:ccs3}
    \end{subfigure}
    \caption{\textcolor{Reviewer12}{A schematic representation of the CCS method in the worst case scenario: when the root $s^*$ happens not to lie between two successive iterations. Because of the high accuracy of the spline $P_k(s)$, consecutive iterations are very close to each other and immediately "land" within a bracketed interval. Within this interval $P_k(s)$ exactly recovers $\VolFracTilde(s)$, and thus the interface position.}}
    \label{fig:ccs}
\end{figure}

\textcolor{Reviewer12}{This very straightforward way of interpolating $\VolFracTilde(s)$ that exactly recovers $\VolFracTilde(s)$ within a bracketed interval is the main contribution of the CCS algorithm, compared to other methods that require complex geometrical parameterizations \citep{Diot2014, Diot2016, Lopez2018, Lopez2019} to achieve the same. Furthermore, directly compared to NCS \citep{Chen2019}, CSS requires overall significantly less iterations, especially at the ends of the search interval, where the volume fraction derivative diminishes.}

\textcolor{Reviewer12}{The two-point root finding approach is novel compared to only using $\VolFracTilde'(s)$} in a Newton iteration by the NCS algorithm of \citet{Chen2019}. \textcolor{Reviewer12}{The number of iterations of the CCS algorithm are not impacted by the way $\VolFracTilde(s),\VolFracTilde'(s)$ are computed. Different geometrical models and algorithms can be used to compute $\VolFracTilde(s),\VolFracTilde'(s)$. As long as $\VolFracTilde(s),\VolFracTilde'(s)$ are computed near machine-epsilon accuracy (e.g. using algorithms from \citep{Lopez2019}, or from \citep{Ahn2007}), the number of iterations performed by CCS will not change. Because the quadratic nature of the intersection area results in \cref{eq:cubicfrac}, extending the two-point CCS approach to include a polynomial interpolation with order $K>3$, by storing data from more than two consecutive iterations brings no benefit.} In fact, increasing the polynomial order beyond $3$ destabilizes the root finding process because higher-order polynomials often have more than a single root $s^*$ in the search interval $(s_a,s_b)$.

\begin{algorithm}[htb]
    \centering
    \caption{\textcolor{Reviewer1}{Consecutive Cubic Spline (CCS) iterative positioning: Part I}}
    \label{alg:ccsI}
    {\footnotesize
        \begin{algorithmic}[1]
            \State \textbf{Input} $\VolFrac_c, \n_c, V_c$  
            \State \textbf{Return}  $s^*, \p(s^*)$
            \State Calculate $\xMin = \xMin(\n_c, V_c)$ using \cref{eq:xmin}.
            \State $\VolFracTilde(\sMin) = 1 - \VolFrac_c,  \VolFracTilde'(\sMin) = 0$ \Comment \Cref{eq:cubicpoly} conditions for $\sMin$
            \State $\VolFracTilde(\sMax) = - \VolFrac_c,\VolFracTilde'(\sMax) = 0$ \Comment \Cref{eq:cubicpoly} conditions for $\sMax$
            \State Interpolate $P_k(s)$ over $(\sMin,\sMax)$ by Hermite cubic spline.
            \State $s_n = root(P_k(s))$ \Comment Initial estimate from a Hermite cubic spline root.
            \State $s_{n-1} = 0, \VolFracTilde_{n-1} = 0, \VolFracTilde'(s_{n-1}) = 0$ \Comment Initialize data from the previous iteration.
            \State $s_a = \sMin, \VolFracTilde(s_a) = \VolFracTilde(\sMin)$ \Comment Data for the root-containing interval $[s_a,s_b]$. 
            \State $s_b = \sMax, \VolFracTilde(s_b) = \VolFracTilde(\sMax)$ \Comment Data for the root-containing interval $[s_a,s_b]$.
            \For{$n = 0, n  < \text{MAX\_ITERATIONS}, n=n+1$}
              \State $\p(s_n) = \xMin + s_n \n_c$ 
              \State $H(s_n) = H(\p(s_n), \n_c)$
              \State $V(s_n), A(s_n) \leftarrow \text{intersect}(H(s_n), V_c)$.  \Comment Truncated volume and \textcolor{Reviewer12}{cap area}.
              \State $\VolFracTilde(s_{n-1}) = \VolFracTilde(s_n)$ \Comment Store old $\VolFracTilde$ before update. 
              \State $\VolFracTilde(s_n) = \dfrac{V(s^n)}{V_c} - \VolFrac_c$. 
              \If{$|\VolFracTilde(s_n)| < \epsilon_P$} \Comment If positioning error is below tolerance. 
                \State \Return $s_n, \p(s_n)$ \Comment Terminate.
              \EndIf
              \State $\VolFracTilde^{'}(s_{n-1}) = \VolFracTilde'(s_n)$  \Comment Store old $\VolFracTilde'$ before update.
              \State $\VolFracTilde'(s_n) = -\dfrac{A(s_n)}{V_c}$.
              \If{$\VolFracTilde(s_a) \cdot \VolFracTilde(s_n) < 0$} \Comment Update $[s_a,s_b]$. 
                \State $s_b = s_n, \VolFracTilde(s_b) = \VolFracTilde(s_n)$
              \ElsIf {$\VolFracTilde(s_n) \cdot \VolFracTilde(s_b) < 0$}
                \State $s_a= s_n, \VolFracTilde(s_a) = \VolFracTilde(s_n)$
              \EndIf
              \algstore{myalg}
        \end{algorithmic}
    }
\end{algorithm}

The first part of the proposed CCS iterative positioning method is shown in \cref{alg:ccsI} and the CCS interpolation part is shown in \cref{alg:ccsII}. The initial \textcolor{Reviewer12}{guess of the CCS method is calculated using the Hermite cubic interpolation over $[s_{min}, s_{max}]$, with boundary conditions given by \cref{eq:cubicpoly} at $s_{min}, s_{max}$, as shown schematically in \cref{fig:ccs1}}. To use the CCS interpolation between consecutive iterations \textcolor{Reviewer12}{in next iterations}, the interval $[s_a, s_b]$ \textcolor{Reviewer12}{is updated in \cref{alg:ccsI} with data from the current iteration $s_n$ ($s_2$ in \cref{fig:ccs1}).} 

In the first iteration, \textcolor{Reviewer12}{there is only a single root candidate available, shown as $s_2$ schematically in \cref{fig:ccs2}}. Hermite cubic spline is \textcolor{Reviewer12}{therefore} interpolated \textcolor{Reviewer12}{in the next step over} \textcolor{Reviewer12}{the initial interval $[s_{min}, s_{max}]$ ($[s_0, s_1]$ in \cref{fig:ccs2})}, \textcolor{Reviewer12}{together with the first root candidate $s_n$ and its value and derivative pair $(\VolFracTilde(s_n),\VolFracTilde'(s_n))$}. 

For all further iterations $(n > 0)$, as shown in \cref{alg:ccsII}, the CCS interpolation with $P_k(s)$ is performed between $[s_a, s_{b}]$ using values and derivatives from previous root candidates. \textcolor{Reviewer12}{If $[s_a,s_b]$ is a subset of a bracketing interval, and the root of $P_k(s)$ is inside $[s_a, s_b]$, the position of the interface (root of $\VolFracTilde(s)$) is exactly equal to the root of $P_k(s)$.} For the results presented in this manuscript, Brent's method \citep{Brent1971}  is used to find the root of the polynomial \textcolor{Reviewer12}{with machine epsilon tolerance}. Finding the root of $P_k(s)$ using the Brent's method costs only a small fraction of overall computational cost and \textcolor{Reviewer12}{the choice of the polynomial root-finding algorithm has no impact on the number of iterations of the CCS method as long as the polynomial root is calculated within a machine-epsilon tolerance}. 
\begin{algorithm}[htb]
    {\footnotesize
    \begin{algorithmic}
    \caption{\textcolor{Reviewer12}{Consecutive Cubic Spline (CCS) iterative positioning: Part II}}
    \label{alg:ccsII}
    \algrestore{myalg}
              \If{$n == 0$} \Comment A single root candidate is available in the first iteration.
                  \State Interpolate $P_k(s)$ over $[s_0, s_n, s_1]$. \Comment $[s_0, s_1]$ is the initial search interval.
                  \State $P_k(s_0) = \VolFracTilde(s_0), P_k(s_1) = \VolFracTilde(s_1), P_k(s_n)  = \VolFracTilde(s_n), P_k'(s_n) = \VolFracTilde'(s_n)$. 
                  \State $s_{n+1} = \text{root}(P_k(s)) \in [s_0, s_1]$ 
              \Else{}
                  \State Interpolate $P_k(s)$ over $[s_a, s_b]$ by Consecutive Cubic Spline.
                  \State $P_k(s_a) = \VolFracTilde(s_{a}), P_k'(s_{a}) = \VolFracTilde'(s_{a}), P_k(s_b) = \VolFracTilde(s_b), P_k'(s_b) = \VolFracTilde'(s_b)$.
                  \State $s_{n-1} = s_n$ \Comment Store the old position before update.
                  \If {$P_k(s_a) \cdot P_k(s_b) < 0$} 
                    \State $s_{n+1} = \text{root}(P_k(s)) \in [s_a, s_b]$. \Comment Compute the CCS root. 
                    \If {$|P_k(s_{n+1})| < \epsilon_P$} \Comment If the polynomial root value is below tolerance.
                        \State \Return $s_{n+1}, \p(s_{n+1})$ \Comment Terminate.
                    \EndIf
                    \If {$n > 2$ and $|\VolFracTilde'(s_a)| < \epsilon$ or $|\VolFracTilde'(s_b)| < \epsilon$} \Comment Non-zero derivative at $s_a$ or $s_b$.
                        \State $s_{n+1} = 0.5 (s_a + s_b)$ \Comment Bisect.
                    \EndIf
                  \Else{}
                    \State $s_{n+1} = s_n - \dfrac{\VolFracTilde(s_n)}{\VolFracTilde'(s_n)}$ \Comment Newton iteration.
                  \EndIf
              \EndIf 
            \EndFor
      \end{algorithmic}
    }
\end{algorithm}

\textcolor{Reviewer12}{An important detail regarding $P_k(s)$ is the accurate calculation \textcolor{Reviewer23}{of} the polynomial coefficients for $P_k(s)$. An error in the calculation of $P_k(s)$ coefficients means $P_k(s)$ cannot exactly interpolate $\VolFracTilde(s)$ in $[s_a, s_b]$, resulting, in turn, with an erroneous interface position, even if the root of $P_k(s)$ is calculated exactly. Parameterizing $s$ such that $s \in [0,1]$, enables a numerically stable way to exactly calculate the coefficients of $P_k(s)$ as}
\textcolor{Reviewer12}{\begin{equation}
    \begin{aligned}
    a_0 & = \VolFracTilde(s_a) \\ 
    a_1 & = -(\VolFracTilde'(s_a)s_a - \VolFracTilde'(s_a)s_b) \\
    a_2 & = -(-2\VolFracTilde'(s_a)s_a + 2\VolFracTilde'(s_a)s_b - \VolFracTilde'(s_b)s_a + \VolFracTilde'(s_b)s_b + 3\VolFracTilde(s_a) - 3\VolFracTilde(s_b)) \\
    a_3 & = -(\VolFracTilde'(s_a)s_a- \VolFracTilde'(s_a)s_b + \VolFracTilde'(s_b)s_a - \VolFracTilde'(s_b)s_b - 2\VolFracTilde(s_a) + 2\VolFracTilde(s_b)).
    \end{aligned}
    \label{eq:excoeffs}
\end{equation}}
\textcolor{Reviewer12}{An unnecessary intersection is removed by line $37$ in \cref{alg:ccsII}, in the case when the root value of $P_k(s)$ falls below the positioning tolerance, because when $s^* \in [s_a,s_b]$, $P_k(s)$ recovers $\VolFracTilde(s)$ exactly.} 

\textcolor{Reviewer12}{In the case when $s^* \notin [s_a, s_b]$, a Newton iteration is used, as shown schematically in \cref{fig:ccs3}. A separate special case of $s^* \notin [s_a, s_b]$, happening over multiple iterations, is caused by the collinearity of the halfspace normal $\n_c$ with the normal vector of \emph{a planar face} of $V_c$. This leads to $\VolFracTilde'(s_b) \ne 0$ ($\VolFracTilde'(s_a) \ne 0$) where $s_{b} = s_{max}$ ($s_{a} = s_{min}$), which is contrary to the boundary conditions \ref{eq:cubicpoly}. This special case is handled in a very straightforward way by CCS, by lines $40-42$ in \cref{alg:ccsII}. In the case where CCS needs more than $2$ iterations and one of the boundary derivatives is zero, this is a possible special case with a falsely assumed zero derivative at one of the interval boundaries. In this case a simple bisection step is performed, because it will force a truncation and thus the calculation of the actual $\VolFracTilde'(s_{n+1})$, replacing the assumed zero derivative at $s_{min}$ or $s_{max}$ with the calculated derivative at $s_{n+1}$.}  

%% file: figures/ccs-iteration1.pdf_tex
\begingroup%
  \makeatletter%
  \providecommand\color[2][]{%
    \errmessage{(Inkscape) Color is used for the text in Inkscape, but the package 'color.sty' is not loaded}%
    \renewcommand\color[2][]{}%
  }%
  \providecommand\transparent[1]{%
    \errmessage{(Inkscape) Transparency is used (non-zero) for the text in Inkscape, but the package 'transparent.sty' is not loaded}%
    \renewcommand\transparent[1]{}%
  }%
  \providecommand\rotatebox[2]{#2}%
  \newcommand*\fsize{\dimexpr\f@size pt\relax}%
  \newcommand*\lineheight[1]{\fontsize{\fsize}{#1\fsize}\selectfont}%
  \ifx\svgwidth\undefined%
    \setlength{\unitlength}{170.07874016bp}%
    \ifx\svgscale\undefined%
      \relax%
    \else%
      \setlength{\unitlength}{\unitlength * \real{\svgscale}}%
    \fi%
  \else%
    \setlength{\unitlength}{\svgwidth}%
  \fi%
  \global\let\svgwidth\undefined%
  \global\let\svgscale\undefined%
  \makeatother%
  \begin{picture}(1,0.83333333)%
    \lineheight{1}%
    \setlength\tabcolsep{0pt}%
    \put(0,0){\includegraphics[width=\unitlength,page=1]{ccs-iteration1.pdf}}%
    \put(0.14861811,0.68890527){\makebox(0,0)[lt]{\lineheight{1.25}\smash{\begin{tabular}[t]{l}$\tilde{\alpha}(s)$\end{tabular}}}}%
    \put(0.79493376,0.05588163){\makebox(0,0)[lt]{\lineheight{1.25}\smash{\begin{tabular}[t]{l}$s$\end{tabular}}}}%
    \put(0,0){\includegraphics[width=\unitlength,page=2]{ccs-iteration1.pdf}}%
    \put(0.64502227,0.12078927){\makebox(0,0)[lt]{\lineheight{1.25}\smash{\begin{tabular}[t]{l}$s_2$\end{tabular}}}}%
    \put(0,0){\includegraphics[width=\unitlength,page=3]{ccs-iteration1.pdf}}%
    \put(0.74355501,0.12090613){\makebox(0,0)[lt]{\lineheight{1.25}\smash{\begin{tabular}[t]{l}$s_1$\end{tabular}}}}%
    \put(0.12145996,0.06308079){\makebox(0,0)[lt]{\lineheight{1.25}\smash{\begin{tabular}[t]{l}$s_0$\end{tabular}}}}%
    \put(0,0){\includegraphics[width=\unitlength,page=4]{ccs-iteration1.pdf}}%
  \end{picture}%
\endgroup%

%% file: figures/ccs-iteration2.pdf_tex
\begingroup%
  \makeatletter%
  \providecommand\color[2][]{%
    \errmessage{(Inkscape) Color is used for the text in Inkscape, but the package 'color.sty' is not loaded}%
    \renewcommand\color[2][]{}%
  }%
  \providecommand\transparent[1]{%
    \errmessage{(Inkscape) Transparency is used (non-zero) for the text in Inkscape, but the package 'transparent.sty' is not loaded}%
    \renewcommand\transparent[1]{}%
  }%
  \providecommand\rotatebox[2]{#2}%
  \newcommand*\fsize{\dimexpr\f@size pt\relax}%
  \newcommand*\lineheight[1]{\fontsize{\fsize}{#1\fsize}\selectfont}%
  \ifx\svgwidth\undefined%
    \setlength{\unitlength}{170.07874016bp}%
    \ifx\svgscale\undefined%
      \relax%
    \else%
      \setlength{\unitlength}{\unitlength * \real{\svgscale}}%
    \fi%
  \else%
    \setlength{\unitlength}{\svgwidth}%
  \fi%
  \global\let\svgwidth\undefined%
  \global\let\svgscale\undefined%
  \makeatother%
  \begin{picture}(1,0.83333333)%
    \lineheight{1}%
    \setlength\tabcolsep{0pt}%
    \put(0,0){\includegraphics[width=\unitlength,page=1]{ccs-iteration2.pdf}}%
    \put(0.14861811,0.68890527){\makebox(0,0)[lt]{\lineheight{1.25}\smash{\begin{tabular}[t]{l}$\tilde{\alpha}(s)$\end{tabular}}}}%
    \put(0.79493376,0.05588163){\makebox(0,0)[lt]{\lineheight{1.25}\smash{\begin{tabular}[t]{l}$s$\end{tabular}}}}%
    \put(0,0){\includegraphics[width=\unitlength,page=2]{ccs-iteration2.pdf}}%
    \put(0.64502227,0.12078927){\makebox(0,0)[lt]{\lineheight{1.25}\smash{\begin{tabular}[t]{l}$s_2$\end{tabular}}}}%
    \put(0,0){\includegraphics[width=\unitlength,page=3]{ccs-iteration2.pdf}}%
    \put(0.74355501,0.12090613){\makebox(0,0)[lt]{\lineheight{1.25}\smash{\begin{tabular}[t]{l}$s_1$\end{tabular}}}}%
    \put(0.12145996,0.06308079){\makebox(0,0)[lt]{\lineheight{1.25}\smash{\begin{tabular}[t]{l}$s_0$\end{tabular}}}}%
    \put(0,0){\includegraphics[width=\unitlength,page=4]{ccs-iteration2.pdf}}%
    \put(0.57215373,0.12173621){\makebox(0,0)[lt]{\lineheight{1.25}\smash{\begin{tabular}[t]{l}$s_3$\end{tabular}}}}%
    \put(0,0){\includegraphics[width=\unitlength,page=5]{ccs-iteration2.pdf}}%
  \end{picture}%
\endgroup%

%% file: figures/ccs-iteration3.pdf_tex
\begingroup%
  \makeatletter%
  \providecommand\color[2][]{%
    \errmessage{(Inkscape) Color is used for the text in Inkscape, but the package 'color.sty' is not loaded}%
    \renewcommand\color[2][]{}%
  }%
  \providecommand\transparent[1]{%
    \errmessage{(Inkscape) Transparency is used (non-zero) for the text in Inkscape, but the package 'transparent.sty' is not loaded}%
    \renewcommand\transparent[1]{}%
  }%
  \providecommand\rotatebox[2]{#2}%
  \newcommand*\fsize{\dimexpr\f@size pt\relax}%
  \newcommand*\lineheight[1]{\fontsize{\fsize}{#1\fsize}\selectfont}%
  \ifx\svgwidth\undefined%
    \setlength{\unitlength}{7.68374777bp}%
    \ifx\svgscale\undefined%
      \relax%
    \else%
      \setlength{\unitlength}{\unitlength * \real{\svgscale}}%
    \fi%
  \else%
    \setlength{\unitlength}{\svgwidth}%
  \fi%
  \global\let\svgwidth\undefined%
  \global\let\svgscale\undefined%
  \makeatother%
  \begin{picture}(1,0.63326562)%
    \lineheight{1}%
    \setlength\tabcolsep{0pt}%
    \put(0,0){\includegraphics[width=\unitlength,page=1]{ccs-iteration3.pdf}}%
    \put(-8.81162311,13.42343381){\makebox(0,0)[lt]{\lineheight{1.25}\smash{\begin{tabular}[t]{l}$\tilde{\alpha}(s)$\end{tabular}}}}%
    \put(0,0){\includegraphics[width=\unitlength,page=2]{ccs-iteration3.pdf}}%
    \put(-9.41276497,-0.42910867){\makebox(0,0)[lt]{\lineheight{1.25}\smash{\begin{tabular}[t]{l}$s_0$\end{tabular}}}}%
    \put(0,0){\includegraphics[width=\unitlength,page=3]{ccs-iteration3.pdf}}%
    \put(0.74442373,0.48441143){\makebox(0,0)[lt]{\lineheight{1.25}\smash{\begin{tabular}[t]{l}$s_3$\end{tabular}}}}%
    \put(0.24370746,0.35544814){\makebox(0,0)[lt]{\lineheight{1.25}\smash{\begin{tabular}[t]{l}$s_4$\end{tabular}}}}%
    \put(0,0){\includegraphics[width=\unitlength,page=4]{ccs-iteration3.pdf}}%
    \put(0.35598302,0.46617286){\makebox(0,0)[lt]{\lineheight{1.25}\smash{\begin{tabular}[t]{l}$s^*$\end{tabular}}}}%
  \end{picture}%
\endgroup%

%% file: sections/results.tex
\section{Results}
\label{sec:res}

\begin{figure}[!htb]
    \captionsetup{justification=centering}
    \footnotesize
    \centering
    \begin{subfigure}[b]{0.33\textwidth}
      \centering
      \def\svgwidth{\columnwidth}
        { 
            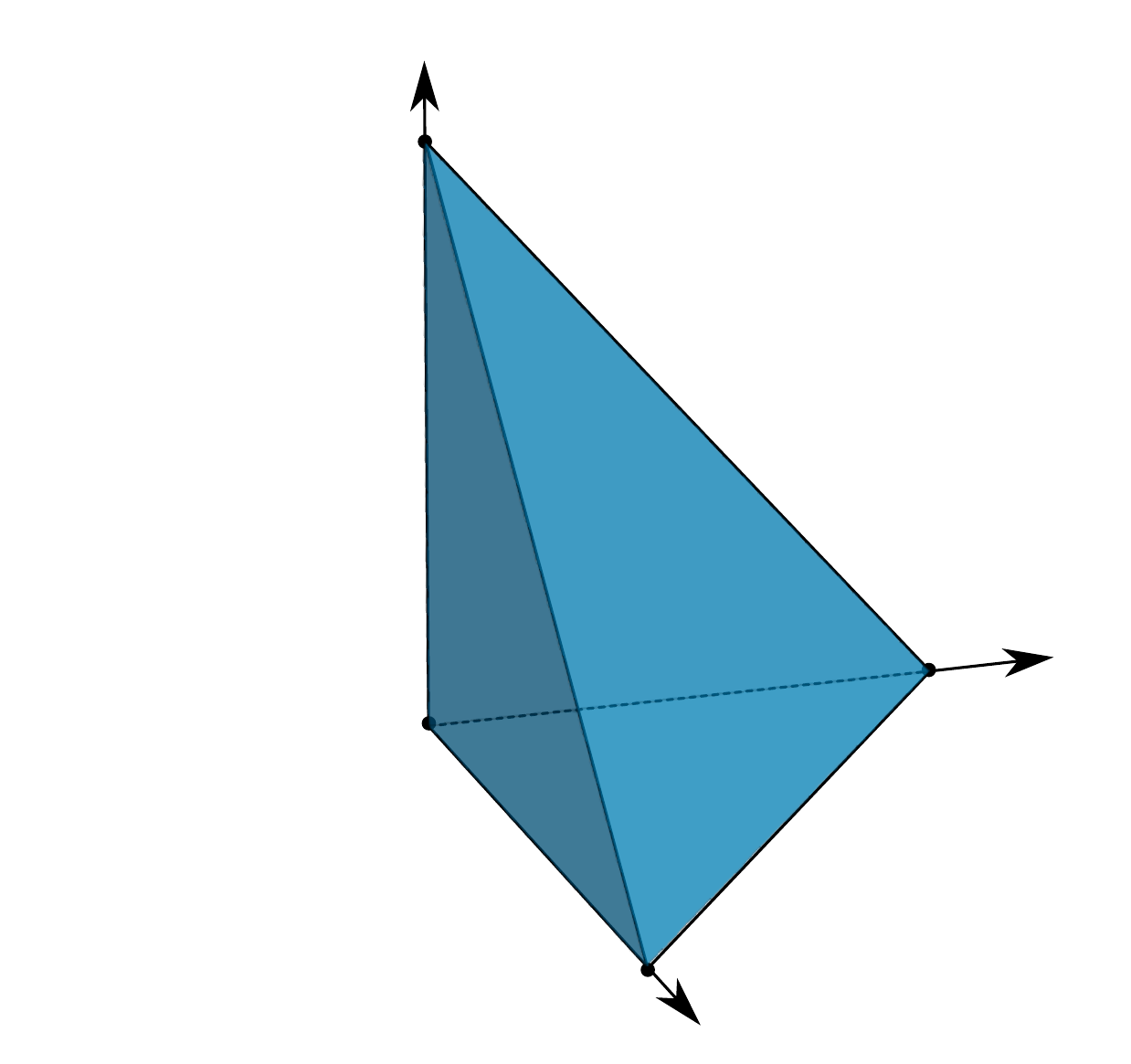 
        }
        \caption{Tetrahedron: TET}
        \label{fig:tetrahedron}
    \end{subfigure}
    \begin{subfigure}[b]{0.33\textwidth}
      \centering
      \def\svgwidth{\columnwidth}
        { 
            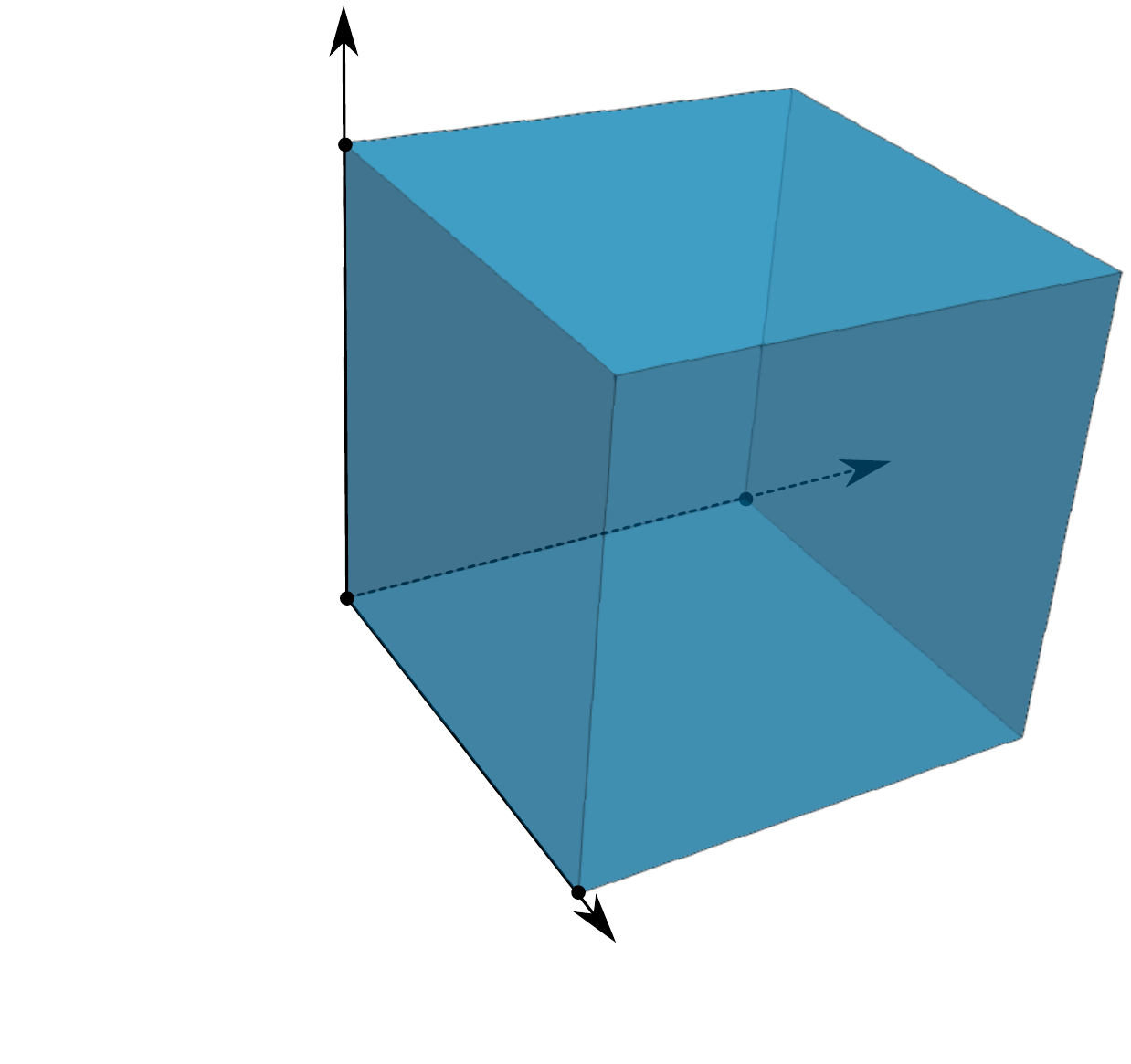 
        }
        \caption{Unit cube: CUBE}
        \label{fig:cube}
    \end{subfigure}
    
    \begin{subfigure}{0.33\textwidth}
      \centering
      \def\svgwidth{\columnwidth}
        { 
            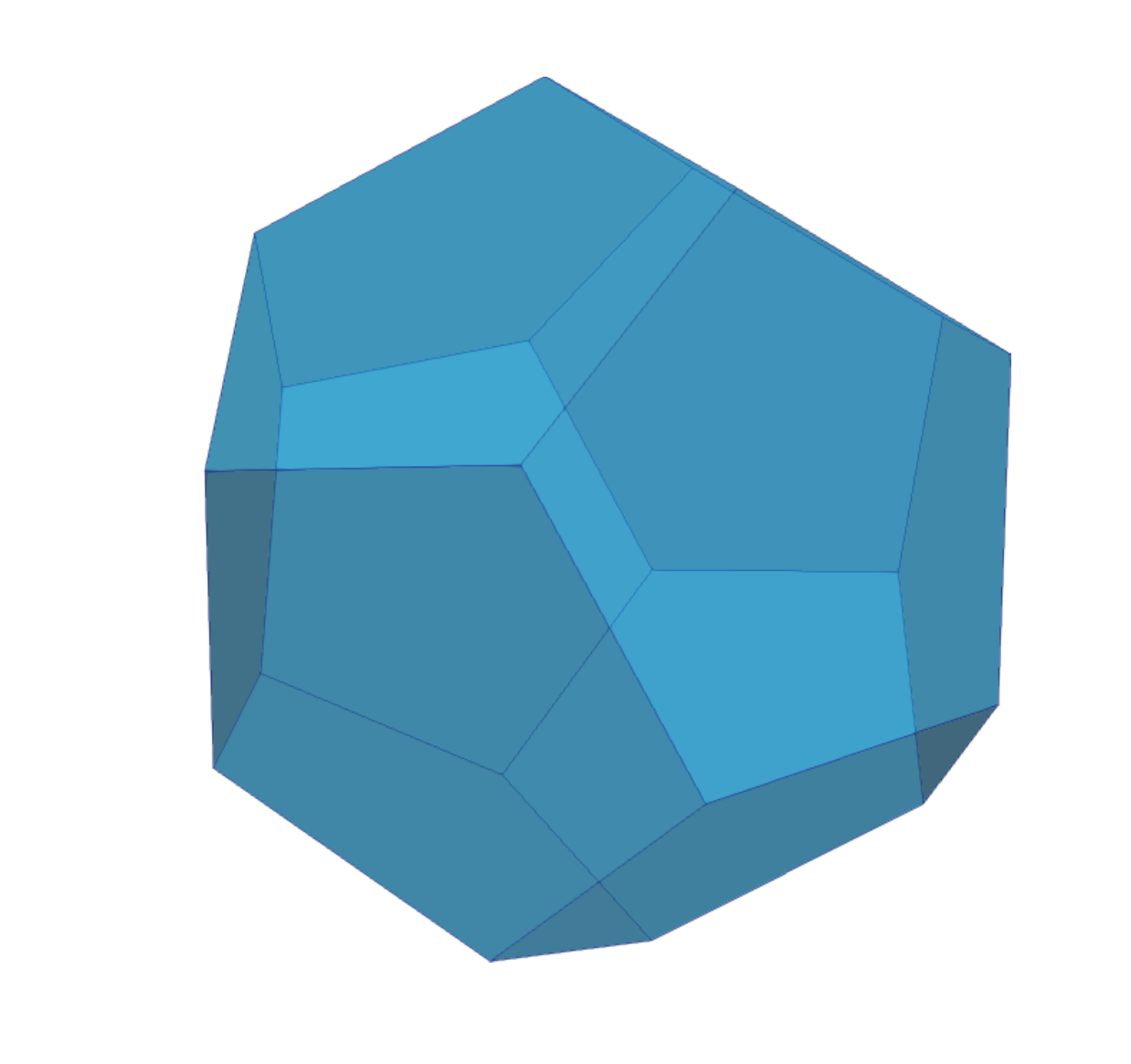 
        }
        \caption{Dodecahedron: DOD}
        \label{fig:dodecahedron}
    \end{subfigure}
    \begin{subfigure}{0.33\textwidth}
      \centering
      \def\svgwidth{\columnwidth}
        { 
            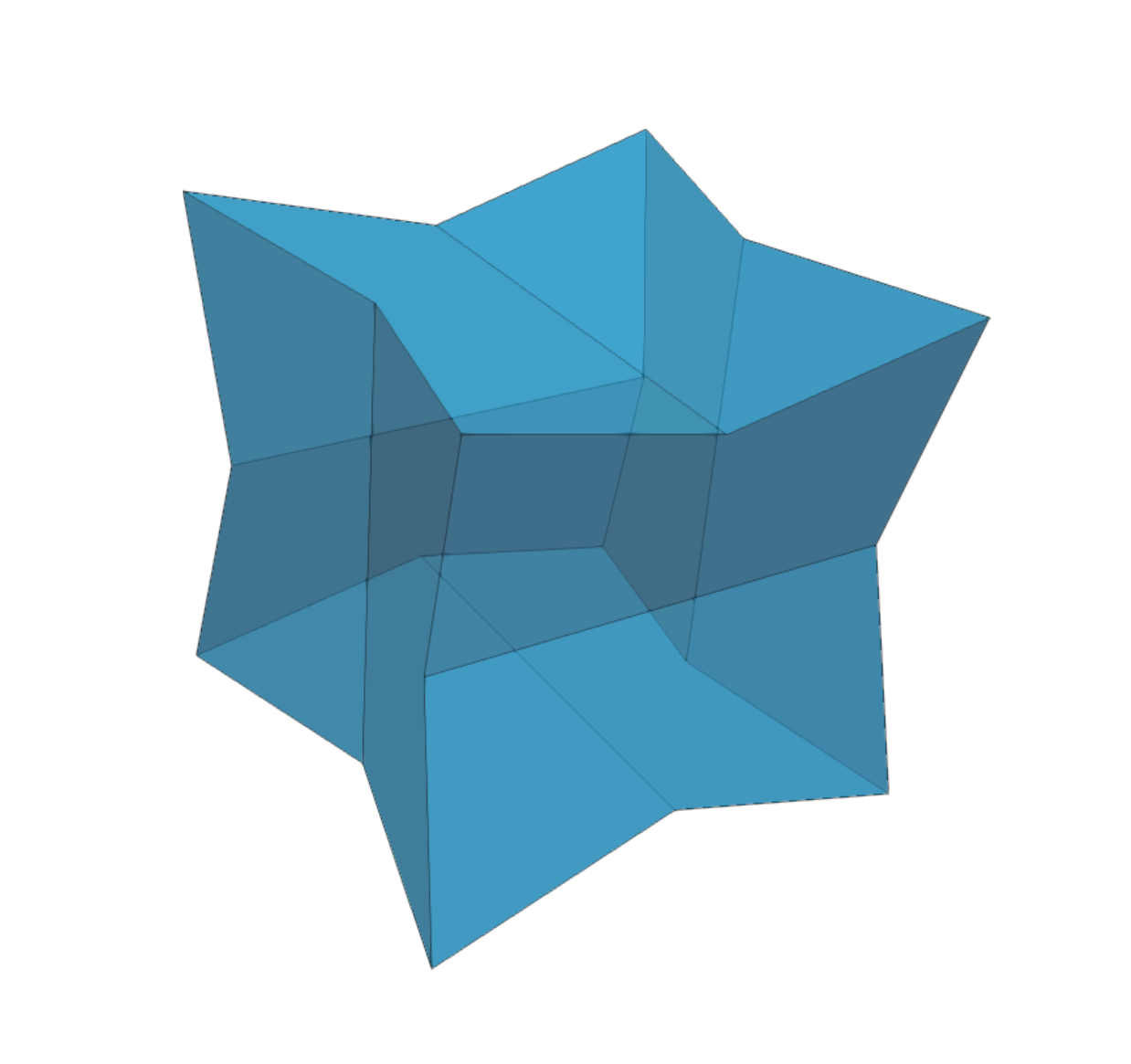 
        }
        \caption{Endo-dodecahedron: ENDO}
        \label{fig:endododecahedron}
    \end{subfigure}

    \begin{subfigure}{0.33\textwidth}
      \centering
      \def\svgwidth{\columnwidth}
        { 
            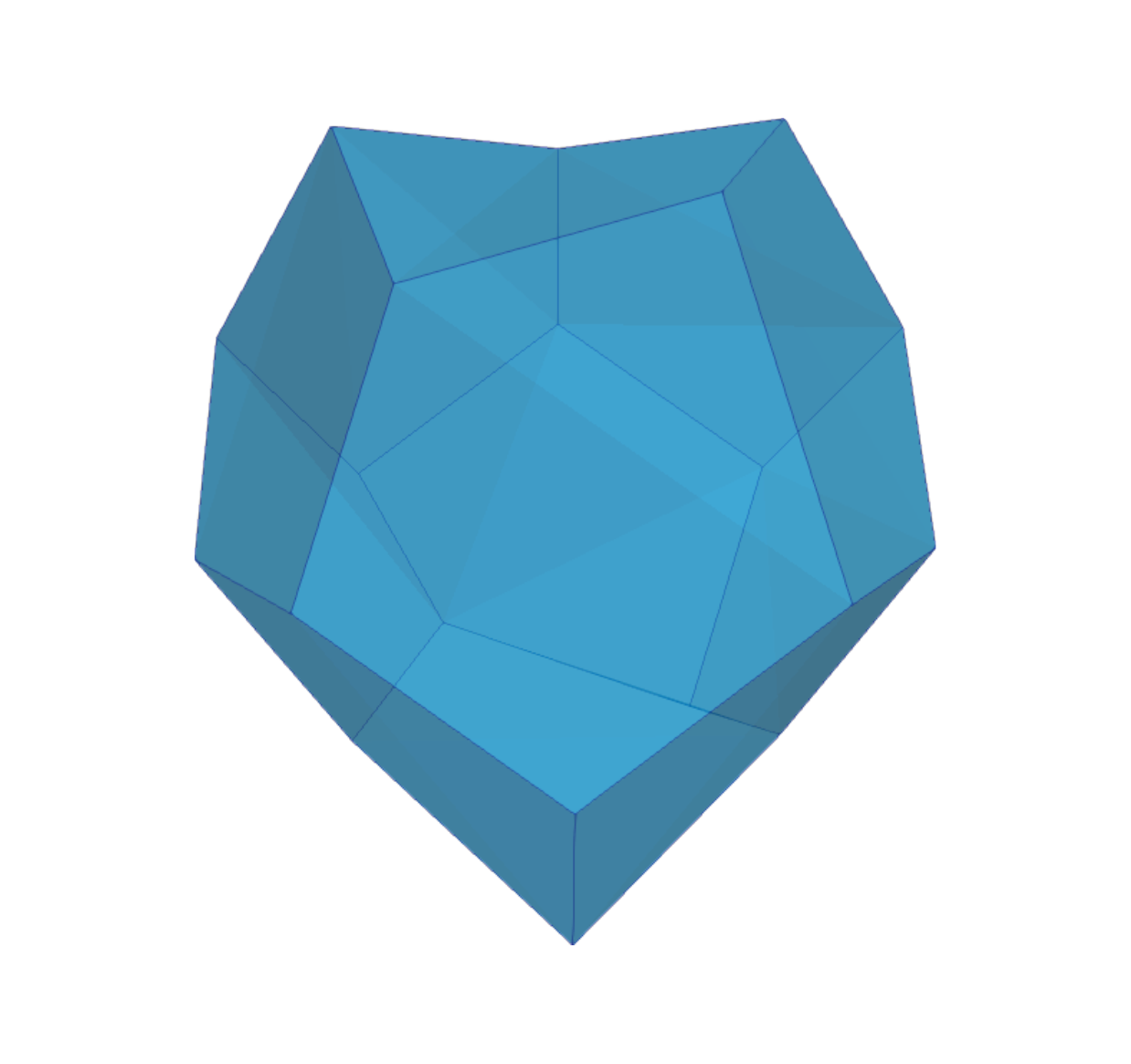 
        }
        \caption{Non-planar dodecahedron: NPDO}
        \label{fig:nonplanardodecahedron}
    \end{subfigure}
    \caption{Test polyhedrons.}
    \label{fig:testpolyhedrons}
\end{figure}

\textcolor{Reviewer2}{The CCS positioning algorithm is tested in \textcolor{Reviewer12}{cells with different shapes}, \textcolor{Reviewer12}{using} challenging volume fraction values and interface orientations, to ensure robust convergence on unstructured meshes. The cells used for testing are shown in \cref{fig:testpolyhedrons}, namely: the tetrahedron (TET), the unit cube (CUBE), the dodecahedron (DOD), the endo-dodecahedron (ENDO) and the non-convex dodecahedron with non-planar faces (NPDO). The endo-dodecahedron represents a model of a non-convex polyhedron that has planar faces. The non-planar dodecahedron is ubiquitous in unstructured polyhedral meshes that are generated by agglomerating tetrahedral cells into polyhedral cells.}

Point coordinates of dodecahedrons from \cref{fig:dodecahedron,fig:endododecahedron,fig:nonplanardodecahedron} are defined using
\begin{align}
    ( \pm 1, \pm 1, \pm 1) + \mathbf{\delta}_{nonplanar}, \\
    (0, \pm(1+h), \pm(1-h^{2})) + \mathbf{\delta}_{nonplanar}, \\ 
    (\pm(1+h), \pm(1-h^{2}), 0) + \mathbf{\delta}_{nonplanar},\\
    (\pm(1-h^{2}), 0, \pm(1+h)) + \mathbf{\delta}_{nonplanar},
\end{align}
where $\mathbf{\delta}_{nonplanar} = (t,t,t)$ is the perturbation vector added to odd points and subtracted from even points in order to cause non-planarity of polyhedron faces. The dodecahedron in \cref{fig:dodecahedron} is computed with $h=0.5(-1 + \sqrt(5)), \mathbf{\delta}_{nonplanar} = \mathbf{0}$, the endo-dodecahedron is computed with $h=-0.25, \mathbf{\delta}_{nonplanar} = \mathbf{0}$, and the non-planar dodecahedron is computed using $h=0.5(-1 + \sqrt(5)), \mathbf{\delta}_{nonplanar} = (0.2,0.2,0.2)$.

The CCS positioning is compared to NCS \citep{Chen2019} using following test parameters 
\textcolor{Reviewer13}{\begin{align}
    \theta     \in \, & (0,\frac{\pi}{N_\theta}, \frac{2\pi}{N_\theta}, \frac{3\pi}{N_\theta}, \dots, \pi) \label{eq:thetaparam} \\
    \phi       \in \, & (0,\frac{\pi}{2N_\theta}, \frac{2\pi}{2N_\theta}, \frac{3\pi}{2N_\theta}, \dots, 2\pi) \label{eq:phiparam} \\
    \VolFrac_c \in \, & (10^{-9},10^{-8},10^{-7},\dots, 10^{-3}, 10^{-3} + \frac{1-2\cdot10^{-3}}{N_\alpha},  \nonumber \\
                   \, & \nonumber 10^{-3} + 2\frac{1-2\cdot10^{-3}}{N_\alpha}, \dots, 1-10^{-3}, \nonumber\\
                   \, & 1-10^{-4},1-10^{-5},1-10^{-6},1-10^{-7},1-10^{-8},1-10^{-9})
    \label{eq:alphaparam} 
\end{align}}
where $(\phi, \theta)$ are the spherical angles that parametrize the interface normal as
\begin{equation}
    \n_c = (\sin(\theta)\cos(\phi), \sin(\theta)\sin(\phi), \cos(\theta)).
    \label{eq:ncphitheta}
\end{equation}

Values of $\VolFrac_c$ in \cref{eq:alphaparam} are based on the reconstruction tolerance \textcolor{Reviewer13}{$\ReconTol = 10^{-9}$}, proposed in \citep{Ahn2008} and used by the author in \citep{Maric2013,Maric2018}. \textcolor{Reviewer1}{It is crucial to include values \emph{boundary volume fraction values} ($\VolFrac_c \approx 0, \VolFrac_c \approx 1$) in the test data, because positioning algorithms that rely on nonzero derivatives fail at the boundaries of the $\VolFrac_c \in [0,1]$ interval. Boundary volume fractions are common in multiphase flow simulations, where the so-called \emph{wisps} (dimensionally unsplit VOF) and \emph{splitting-errors} (dimensionally split VOF) introduce small variations in $\VolFrac_c$ in cells that are otherwise either full ($\VolFrac_c=1$) or empty $(\VolFrac_c=0)$. Since the derivatives of $\VolFrac_c(s)$ vanish for these volume fraction values (cf. \cref{fig:pos-param}), a positioning algorithm that does not take into account diminishing derivatives will experience either slower convergence, or in the worst case, divergence.} \textcolor{Reviewer1}{The positioning condition is based on the positioning tolerance \textcolor{Reviewer13}{$\epsilon_P < 10^{-12}$}, also used by \citet{Lopez2018}: this forces the CCS and NCS iterative algorithms to perform one more iteration, which further reduces $\epsilon_P$ below machine epsilon for the vast majority of tests. \textcolor{Reviewer23}{The subdivision of the polar angle intervals in \cref{eq:alphaparam} is done for the test cases using $N_\theta=40$. A more detailed interval subdivision is not necessary, because the results demonstrate uniform distributions of both the CPU time, and the number of iterations, with respect to the polar angles.} Furthermore, as the tests results confirm, there is no need to use many $\VolFrac_c$ values between \textcolor{Reviewer13}{$[10^{-3},1-10^{-3}]$}, as the average iterations do not vary significantly within this interval. That is why $N_\alpha=50$ was used for the tests presented here.} 
f

\begin{figure}[!htb]
    \centering
    \begin{subfigure}{\textwidth}
        \centering
        \includegraphics[]{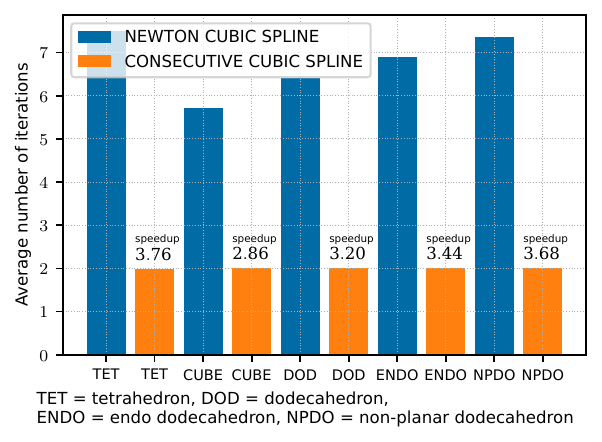}
        \caption{Comparison of average iterations for each cell shape.}
        \label{fig:totaliterations}
    \end{subfigure}
    \begin{subfigure}{\textwidth}
        \centering
            \includegraphics[]{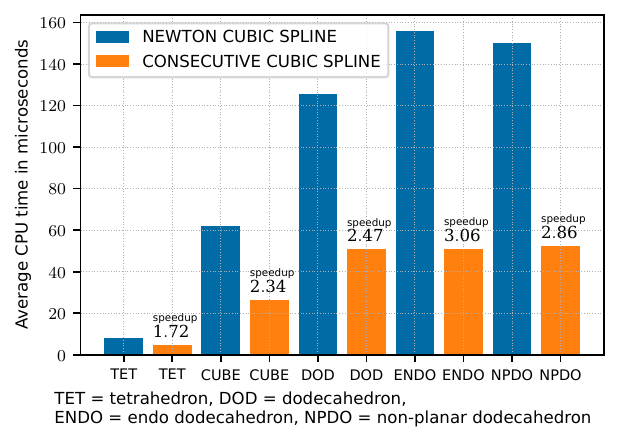}
        \caption{Comparison of average CPU time for each cell shape.}
        \label{fig:totalcputime}
    \end{subfigure}
    \caption{Speedup achieved by the CCS algorithm compared to NCS \citep{Chen2019}.} 
\end{figure}

\textcolor{Reviewer2}{First let us consider the overall results of the CCS algorithm, compared with the NCS algorithm \citep{Chen2019}.} The increase in efficiency in terms of the average number of iterations is presented in \cref{fig:totaliterations}, and the decrease in the average CPU time is shown for different cell shapes in \cref{fig:totalcputime}. An iteration is defined in this context as the calculation of the next interface position $s^{n+1}$. Timing was performed on an architecture given by \cref{tab:testingarch}, and the \textcolor{Reviewer1}{measurements were performed using the \texttt{chrono} C++ standard template library. The same tests were performed on a High-Performance Computing cluster without a significant difference in reported results.} \textcolor{Reviewer12}{A Singularity image \citep{Kurtzer2017}} that contains the source code and the computing environment is publicly available \citep{Singularity} and can be used to easily reproduce the results across different platforms. Results presented in this section using the computing architecture from \cref{tab:testingarch} are also publicly available, together with the source code and binary executables \citep{CodeAndResults}\footnote{The source code is developed at \url{https://git.rwth-aachen.de/leia/geophase}}.

\begin{table}[h]
  \begin{center}
      {\scriptsize
        \input{figures/architectures.tex}
      }
  \end{center}
  \caption{Computer architecture used for testing.}
  \label{tab:testingarch}
\end{table}

\noindent\textcolor{Reviewer2}{The results presented in \cref{fig:totaliterations} show that CCS algorithm requires approximately $2.9 - 3.7$ times less iterations than the NCS algorithm for different polyhedron shapes. In terms of the CPU time shown in \cref{fig:totalcputime}, the CCS algorithm is approximately $1.7 - 3.0$ times faster than the NCS algorithm for different polyhedrons. The stabilized secant / bisection of \citet{Ahn2008} and Brent's method were not used in this comparison, because \citet{Chen2019} already show that those algorithms are significantly outperformed by the NCS algorithm.}

\begin{figure}[!htb]
    \centering
    \begin{subfigure}{0.49\textwidth}
        \includegraphics{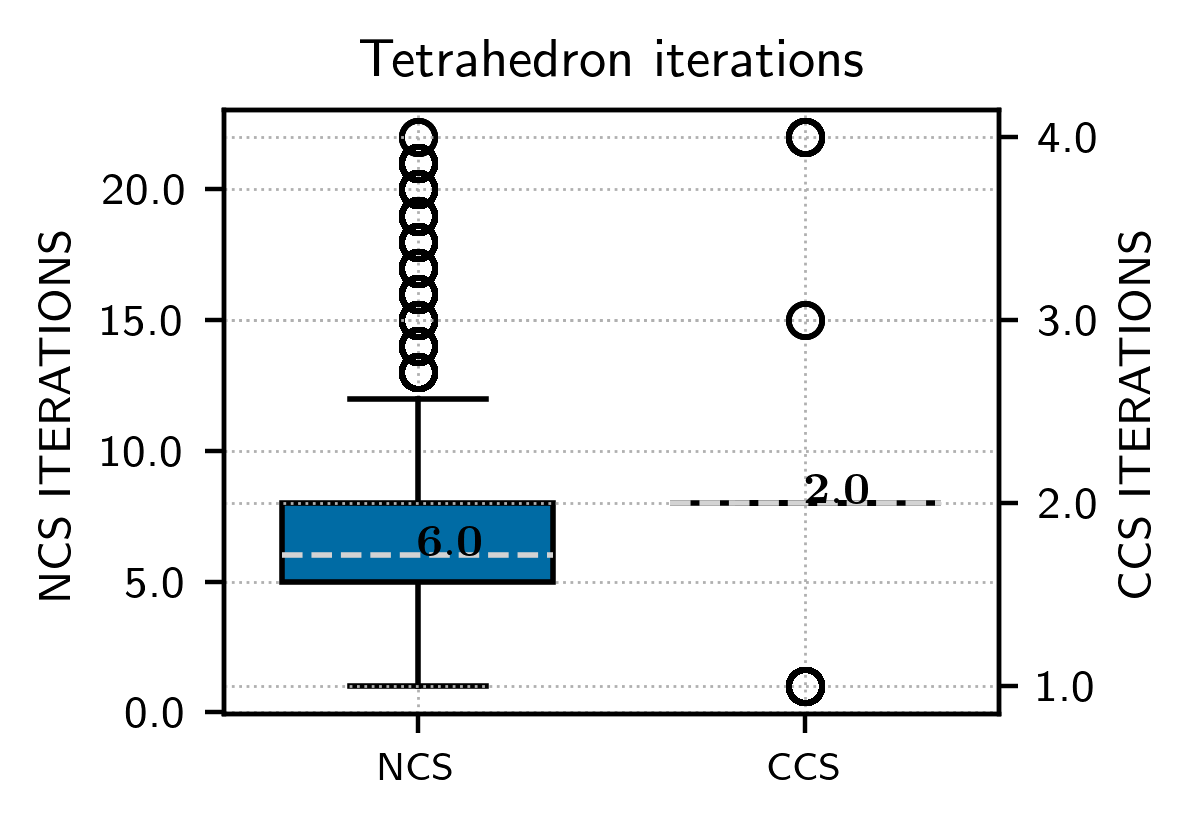}
    \end{subfigure}
    \begin{subfigure}{0.49\textwidth}
        \includegraphics{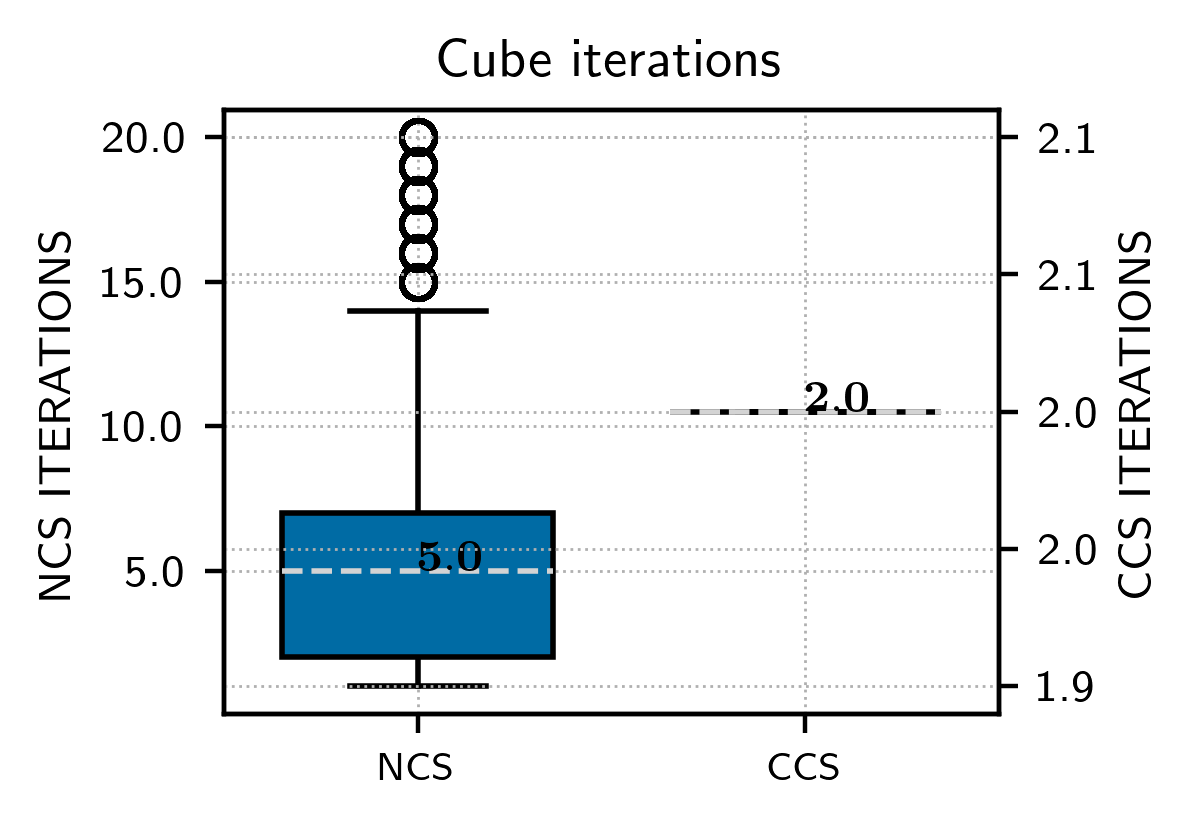}
    \end{subfigure}
    \begin{subfigure}{0.49\textwidth}
        \includegraphics{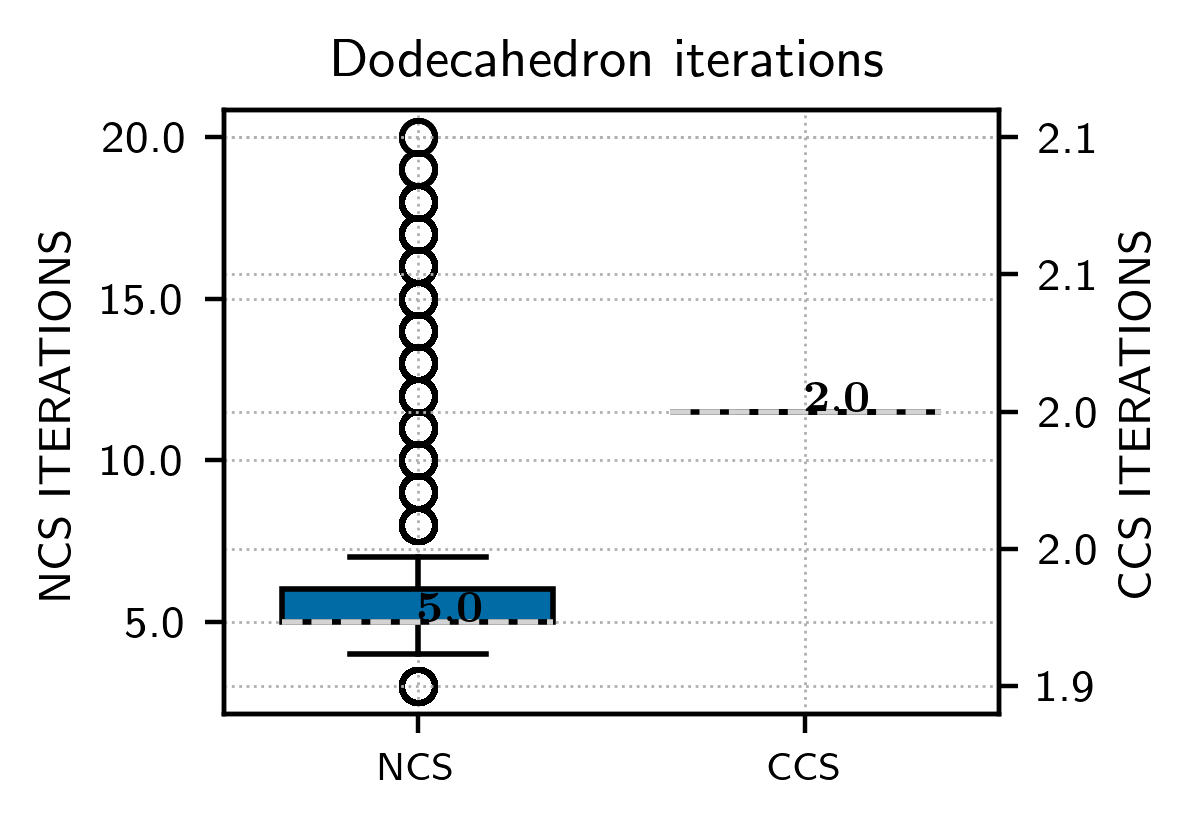}
    \end{subfigure}
    \begin{subfigure}{0.49\textwidth}
        \includegraphics{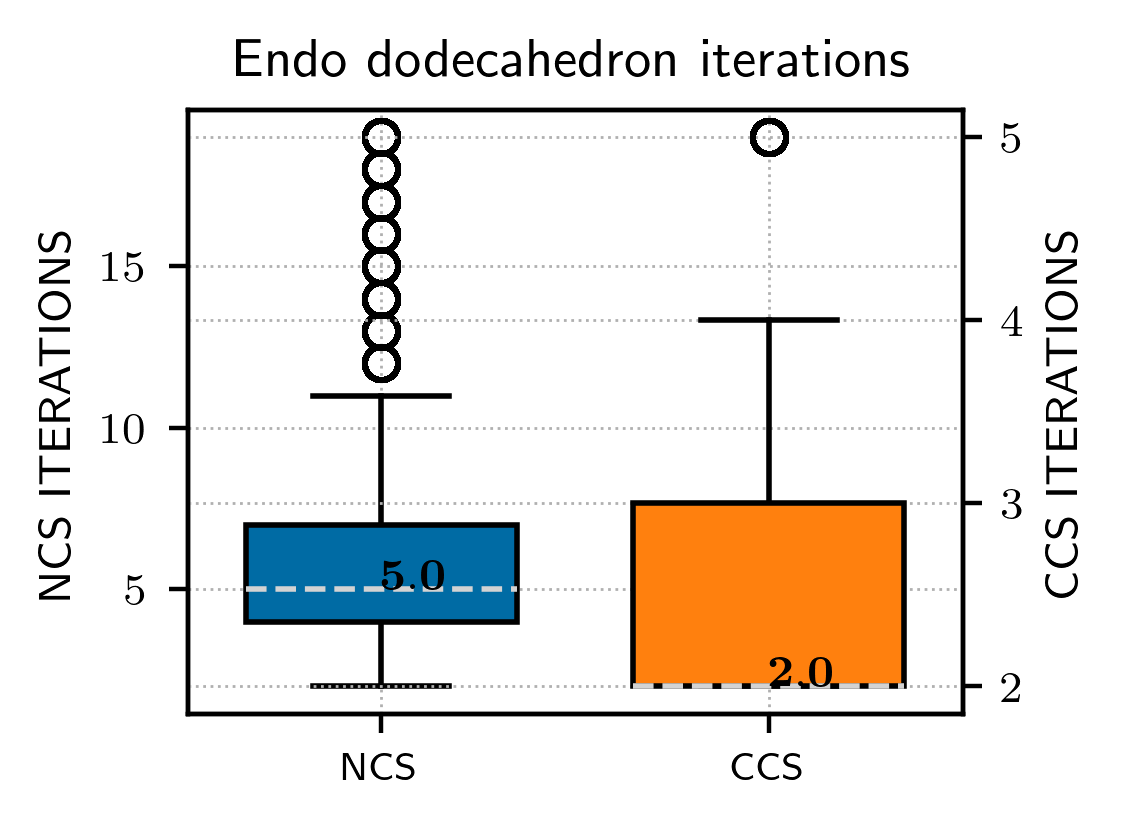}
    \end{subfigure}
    \begin{subfigure}{0.49\textwidth}
        \includegraphics{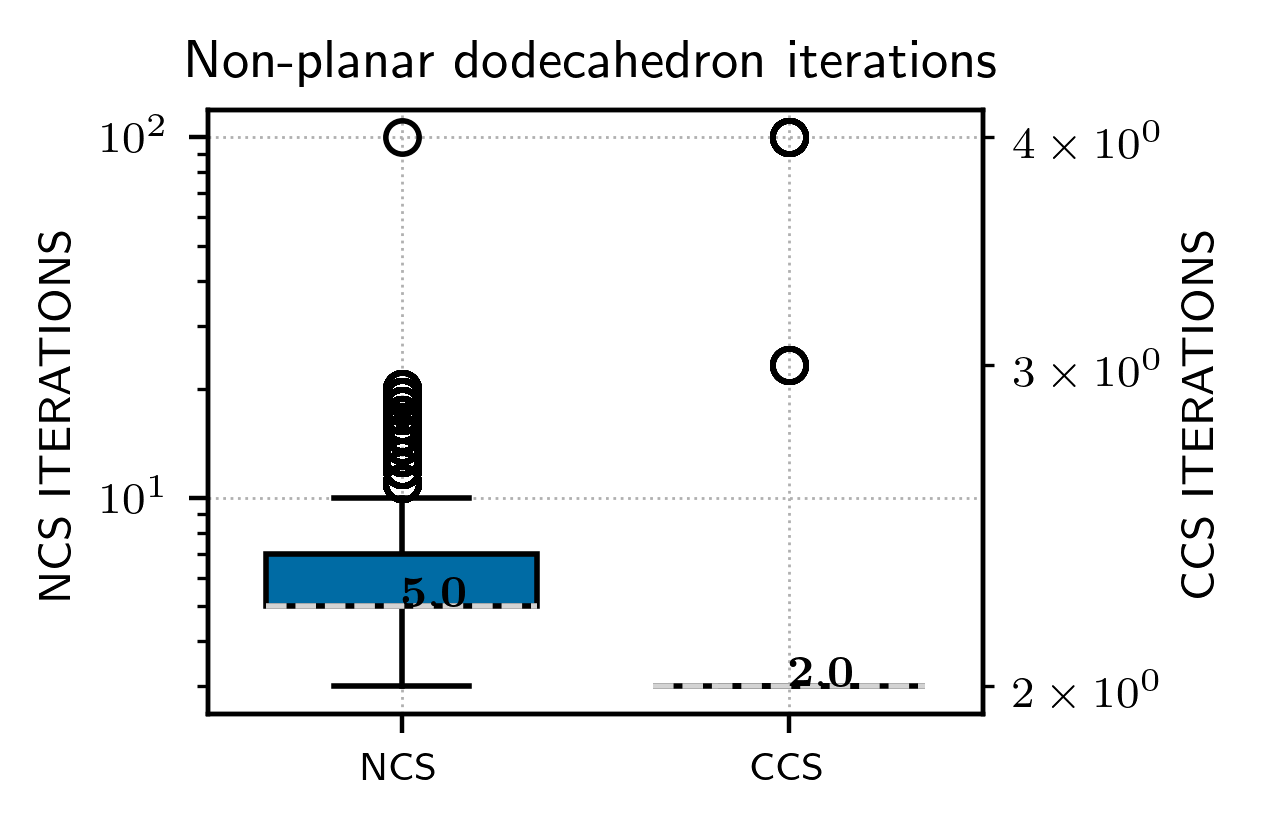}
    \end{subfigure}
    \caption{Iteration distributions of the NCS and CCS algorithms for test polyhedrons.}
    \label{fig:iterboxplots}
\end{figure}

The reduction of \textcolor{Reviewer22}{iterations} achieved by the CCS algorithm is reflected in \cref{fig:iterboxplots} in the iteration distribution for each test polyhedron, when compared to the NCS algorithm. \textcolor{Reviewer12}{Only $2$} iterations are required on average to position the interface within test polyhedrons. Results shown in \cref{fig:totaliterations,fig:totalcputime,fig:iterboxplots} confirm that the CCS algorithm significantly improves the computational efficiency of the interface positioning algorithm on polyhedrons, including star-shaped non-convex polyhedra with non-planar faces. 

\textcolor{Reviewer1}{\citet{Chen2019} use the NCS algorithm in combination with the Moment-of-Fluid (MOF) method and extend the interface positioning problem with the interface orientation angles $\theta,\phi$ that define the interface orientation vector in \cref{eq:ncphitheta}. Using the implicit formulation for the interface plane $n_{c,x} x + n_{c,y}y + n_{c,z}z + d = 0$, the interface position $d$ is defined by \citet{Chen2019} as a multivariate function $d := d(\VolFrac_c,\theta,\phi)$. \citet{Chen2019} then extend the multivariate function $d$ into Taylor series which results in a prediction of the interface position in the new iteration based on $\partial_\theta d$ and $\partial_\phi d$. Consequently, when the information about the changing interface orientation is available from an algorithm that tries to improve the orientation (e.g. MOF), it is possible to find the position of the interface faster. However, it is crucial to note that the orientation information does not impact the partial derivative $\partial_{\VolFrac_c} d$ and the NCS algorithm is used in \citet{Chen2019} independently of the interface orientation prediction. The CCS positioning algorithm improves significantly the estimation of $\partial_{\VolFrac_c} d$ by \textcolor{Reviewer12}{the CCS polynomial interpolation} of $\VolFrac(s)$. \citet[Figure 12]{Chen2019} report an average of $4$ iterations when the orientation information is given by $N_\theta=100 (N_\phi=200)$, and approximately $3$ iterations when the orientation information is given by $N_\theta=1000 (N_\phi=2000)$ for a test case with  \textcolor{Reviewer13}{$\VolFrac_c \in [10^{-3}, 1-10^{-3}]$}.}

\textcolor{Reviewer12}{The CCS algorithm positions the interface with only} \textcolor{Reviewer12}{$2$} average iterations without \textcolor{Reviewer12}{relying on the interface orientation information} (cf. \cref{fig:iterboxplots}), for different cell shapes, and with a more challenging set of $\VolFrac_c$ values given by \cref{eq:alphaparam}. This is confirmed by \citet[Figure 13]{Chen2019}, with more challenging values \textcolor{Reviewer13}{$\VolFrac_c \in [10^{-9},1-10^{-9}]$}, where the average iterations are reported as follows: NCS 13.53 average iterations, predicted NCS ($N_\theta=100$) 6.57 average iterations, and predicted NCS ($N_\theta=1000$) 2.89 average iterations. Therefore, it takes the orientation prediction in \citep{Chen2019} $1000$ subdivisions of the interval $\theta \in [0,\pi]$ to provide enough information to the Taylor series expansion in order to reach the average number of iterations given by the CCS that does not require this information. Because the MOF method requires significantly less than \textcolor{Reviewer13}{$1000$ iterations} to improve the interface orientation, \citet{Chen2019} report an overall increase of $60-66\%$ in efficiency in terms of the average number of iterations. \textcolor{Reviewer12}{It is relevant to note that the implementation of NCS used here for comparison is numerically unstable: a test configuration for the non-planar dodechedron in \cref{fig:iterboxplots} causes the NCS not to converge even after 100 iterations.} 

\textcolor{Reviewer1}{The CIBRAVE method by \citet{Lopez2016,Lopez2018} uses Interpolation Bracketing based on linear interpolation of the bracketing volume, and an explicit (analytical) calculation of the interface position within the bracketed interval. \citet[Fig. 14]{Lopez2016} report a $\log_2(log_2(I_p))$ complexity for the \emph{average number of volume truncation operations} for the CIBRAVE method, where $I_p$ is the number of vertices of the polyhedron. This results in $~2.6$ bracketing truncations on average for a dodecahedron. However, unlike the iterative algorithms, bracketing algorithms \textcolor{Reviewer12}{additionally} require the evaluation of \textcolor{Reviewer12}{a geometrically parameterized} explicit function to position the interface, and this comes at a cost (cf. \citep[Fig 15]{Lopez2016}). \citet{Lopez2016} state that the cost of the coefficient calculation for explicit positioning function is "\emph{around 1.7 times the CPU time needed to make a truncation operation}" \citep{Lopez2016}. 
\citet{Lopez2016} use the test parameters from \citet{Diot2016}, namely \textcolor{Reviewer13}{$\VolFrac \in [10^{-3},1-10^{-3}]$}, which is not as challenging as volume fraction values used here for the CCS method and by \citet[Figure 13]{Chen2019} for the NCS algorithm, so the results are difficult to compare directly. Even though \citet{Lopez2016} also use different cell shapes, the average number of iterations of CCS is $2$, which is comparable to CIBRAVE, while being significantly easier to implement.}

\begin{figure}[htb]
    \includegraphics[scale=0.85]{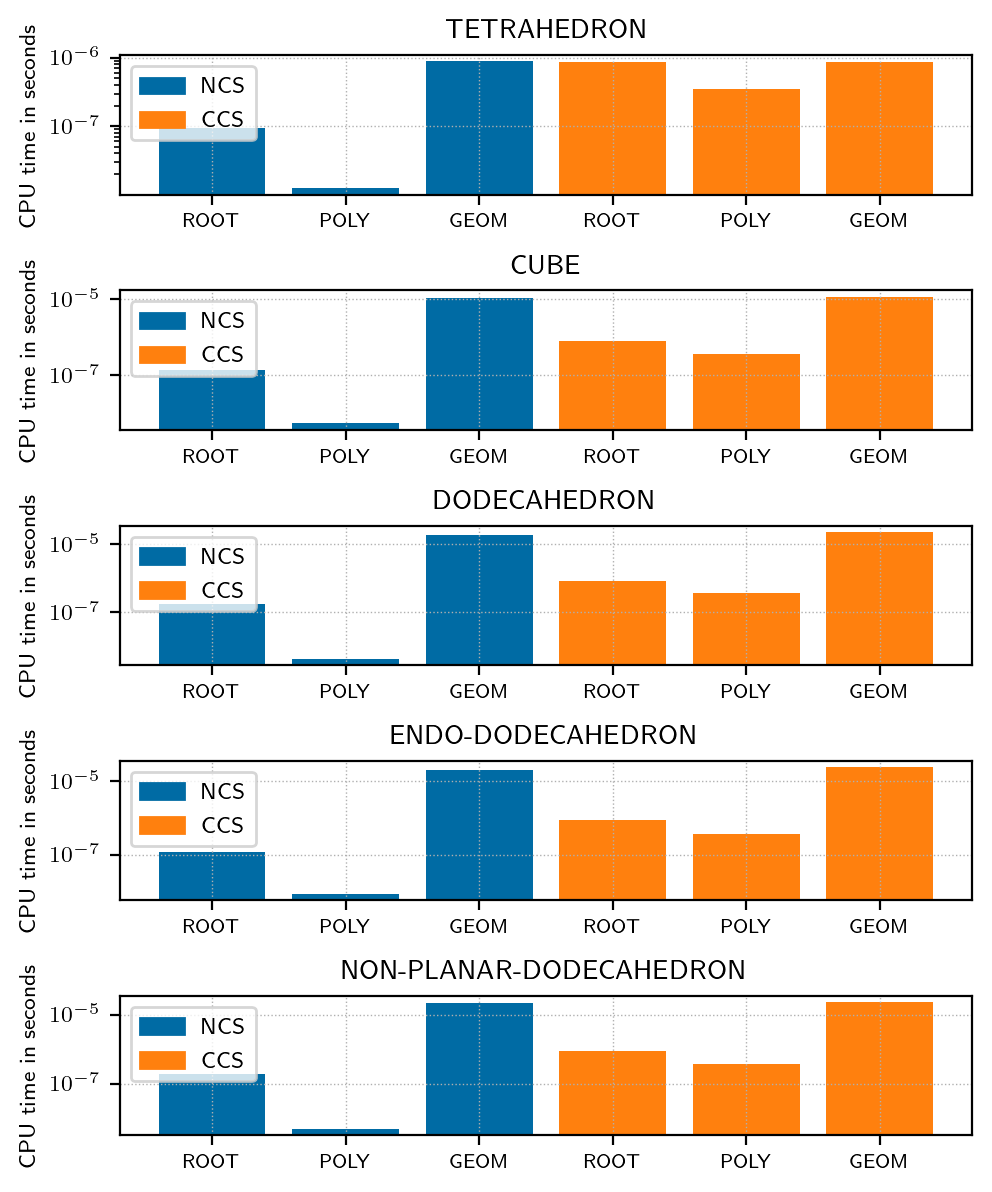}
    \caption{CPU time distribution between the sub-algorithms of the NCS and CCS algorithm.}
    \label{fig:cpudistribution}
\end{figure}

\begin{figure}[htb]
    \centering
    \begin{subfigure}{0.4\textwidth}
        \centering
        \includegraphics[width=\textwidth]{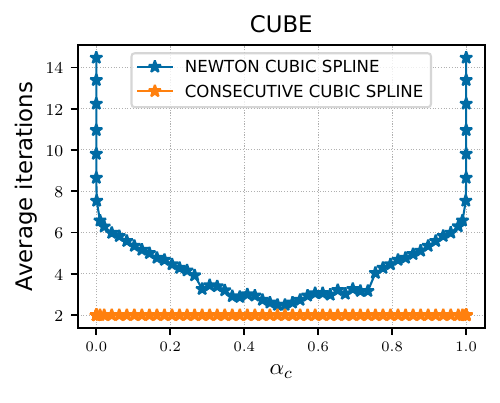}
    \end{subfigure}
    \centering
    \begin{subfigure}{0.4\textwidth}
        \centering
        \includegraphics[width=\textwidth]{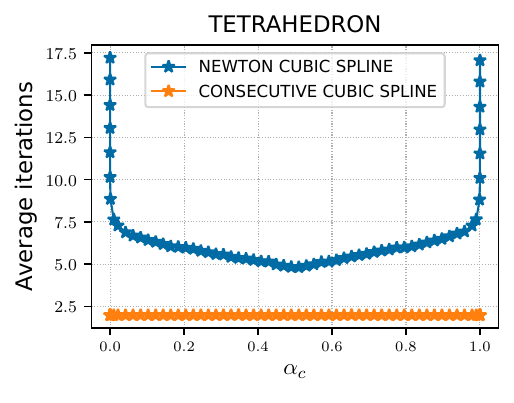}
    \end{subfigure}
    \begin{subfigure}{0.4\textwidth}
        \centering
        \includegraphics[width=\textwidth]{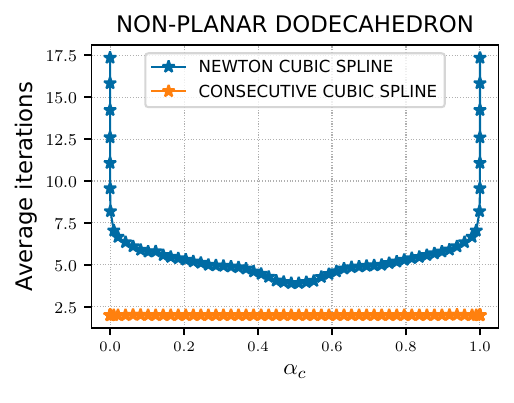}
    \end{subfigure}
    \begin{subfigure}{0.4\textwidth}
        \centering
        \includegraphics[width=\textwidth]{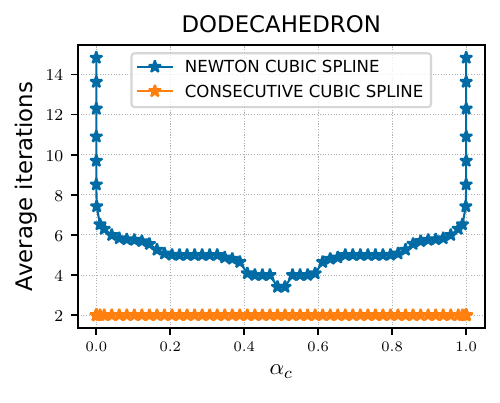}
    \end{subfigure}
    \caption{Average number of iterations depending on given volume fraction $\VolFrac_c$.}
    \label{fig:itervolfrac}
\end{figure}

\textcolor{Reviewer1}{The distribution of the CPU time between sub-algorithms of the CCS method is shown in \cref{fig:cpudistribution}. In \cref{fig:cpudistribution} the CPU time distribution is reported for every test polyhedron and a comparison between the NCS algorithm (blue color) and the CCS algorithm (orange color) is given. The CPU time is distributed between the numerical root finding for the Consecutive Cubic Spline (ROOT), the Consecutive Cubic Spline polynomial interpolation (POLY) and the geometrical volume truncation operation (GEOM). As expected, most of the computational cost is caused by the volume truncation operation, which is why the reduction in the number of iterations in \cref{fig:totaliterations} correlates well with the speedup in terms of CPU time in \cref{fig:totalcputime}. The GEOM CPU time would need to be reduced by approximately \textcolor{Reviewer12}{an order} of magnitude for some polyhedrons in \cref{fig:cpudistribution} to reach the absolute level of ROOT and POLY times and thus impact the relative increase in computational efficiency. \textcolor{Reviewer12}{For the tetrahedron, this is not relevant, as the average positioning CPU time is only $4.5$ microseconds.} \citet{Lopez2019} extend the volume truncation algorithm used by CIBRAVE \citep{Lopez2016,Lopez2018} to non-convex polyhedrons with the aim to maintain the same computational efficiency. \citet{Lopez2019} compare the computational efficiency of the CIBRAVE method with and without their novel truncation using the Brent's method as the basis for comparison. An order of magnitude faster positioning is achieved with CIBRAVE and the new truncation, and four times faster when with CIBRAVE and tetrahedral decomposition. The CCS algorithm would benefit in the absolute CPU time by introducing a faster truncation algorithm (e.g. the one proposed by \citet{Lopez2018,Lopez2019}), however this does not change its relative speedup, resulting from the reduction of the number of truncations. \textcolor{Reviewer12}{The truncation of a cell in an unstructured mesh is closely related to the data structure used to implement the unstructured mesh. To reduce the CPU time required for the interface positioning further, by introducing a new truncation algorithm, existing mesh data structures (and respective cell models) should be used for the truncation operation. If a different cell data structure from the one in the unstructured mesh is used, the geometrical model of the cell must be translated into the model (data structure) that can be used by the positioning algorithm. This model translation incurs additional memory operations that are often computationally expensive and cache-inefficient, because they require access to unstructured data from the main memory. Additional calculations are also required, if an alternative connectivity of cell faces and edges is required by the positioning model, compared to what is available in the mesh. The CCS significantly increases the computational efficiency of interface positioning in a very straightforward way, irrespective of the available truncation algorithm / cell model, and can be readily used in an existing geometrical VOF code, without modification.}}

\textcolor{Reviewer1}{Finally, the reason behind the increased computational efficiency reported in \cref{fig:totaliterations,fig:totalcputime,fig:iterboxplots} is shown in \cref{fig:itervolfrac}. The CCS algorithm outperforms the NCS algorithm \citep{Chen2019} across the interval $\VolFrac_c \in [\epsilon_R, 1-\epsilon_R]$. Additionally, at the boundaries of the interval $[\epsilon_R, 1-\epsilon_R]$, the CCS algorithm achieves a significant reduction in the average number of iterations even when \textcolor{Reviewer13}{$\epsilon_R=10^{-9}$}, which has not yet been reported for other contemporary positioning algorithms, to the best of the author's knowledge.} 

%% file: figures/test-tetrahedron.pdf_tex
\begingroup%
  \makeatletter%
  \providecommand\color[2][]{%
    \errmessage{(Inkscape) Color is used for the text in Inkscape, but the package 'color.sty' is not loaded}%
    \renewcommand\color[2][]{}%
  }%
  \providecommand\transparent[1]{%
    \errmessage{(Inkscape) Transparency is used (non-zero) for the text in Inkscape, but the package 'transparent.sty' is not loaded}%
    \renewcommand\transparent[1]{}%
  }%
  \providecommand\rotatebox[2]{#2}%
  \newcommand*\fsize{\dimexpr\f@size pt\relax}%
  \newcommand*\lineheight[1]{\fontsize{\fsize}{#1\fsize}\selectfont}%
  \ifx\svgwidth\undefined%
    \setlength{\unitlength}{594bp}%
    \ifx\svgscale\undefined%
      \relax%
    \else%
      \setlength{\unitlength}{\unitlength * \real{\svgscale}}%
    \fi%
  \else%
    \setlength{\unitlength}{\svgwidth}%
  \fi%
  \global\let\svgwidth\undefined%
  \global\let\svgscale\undefined%
  \makeatother%
  \begin{picture}(1,0.94065657)%
    \lineheight{1}%
    \setlength\tabcolsep{0pt}%
    \put(0.17529333,0.28184841){\color[rgb]{0,0,0}\makebox(0,0)[lt]{\lineheight{1.25}\smash{\begin{tabular}[t]{l}$(0,0,0)$\end{tabular}}}}%
    \put(0.60451477,0.07848505){\color[rgb]{0,0,0}\makebox(0,0)[lt]{\lineheight{1.25}\smash{\begin{tabular}[t]{l}$(1,0,0)$\end{tabular}}}}%
    \put(0.82160649,0.39790186){\color[rgb]{0,0,0}\makebox(0,0)[lt]{\lineheight{1.25}\smash{\begin{tabular}[t]{l}$(0,1,0)$\end{tabular}}}}%
    \put(0.39942859,0.82472406){\color[rgb]{0,0,0}\makebox(0,0)[lt]{\lineheight{1.25}\smash{\begin{tabular}[t]{l}$(0,0,1)$\end{tabular}}}}%
    \put(0,0){\includegraphics[width=\unitlength,page=1]{test-tetrahedron.pdf}}%
  \end{picture}%
\endgroup%

%% file: figures/test-cube.pdf_tex
\begingroup%
  \makeatletter%
  \providecommand\color[2][]{%
    \errmessage{(Inkscape) Color is used for the text in Inkscape, but the package 'color.sty' is not loaded}%
    \renewcommand\color[2][]{}%
  }%
  \providecommand\transparent[1]{%
    \errmessage{(Inkscape) Transparency is used (non-zero) for the text in Inkscape, but the package 'transparent.sty' is not loaded}%
    \renewcommand\transparent[1]{}%
  }%
  \providecommand\rotatebox[2]{#2}%
  \newcommand*\fsize{\dimexpr\f@size pt\relax}%
  \newcommand*\lineheight[1]{\fontsize{\fsize}{#1\fsize}\selectfont}%
  \ifx\svgwidth\undefined%
    \setlength{\unitlength}{594bp}%
    \ifx\svgscale\undefined%
      \relax%
    \else%
      \setlength{\unitlength}{\unitlength * \real{\svgscale}}%
    \fi%
  \else%
    \setlength{\unitlength}{\svgwidth}%
  \fi%
  \global\let\svgwidth\undefined%
  \global\let\svgscale\undefined%
  \makeatother%
  \begin{picture}(1,0.94065657)%
    \lineheight{1}%
    \setlength\tabcolsep{0pt}%
    \put(0.55512619,0.10281441){\color[rgb]{0,0,0}\makebox(0,0)[lt]{\lineheight{1.25}\smash{\begin{tabular}[t]{l}$(1,0,0)$\end{tabular}}}}%
    \put(0.68482948,0.46104897){\color[rgb]{0,0,0}\makebox(0,0)[lt]{\lineheight{1.25}\smash{\begin{tabular}[t]{l}$(0,1,0)$\end{tabular}}}}%
    \put(0.11039458,0.38693927){\color[rgb]{0,0,0}\makebox(0,0)[lt]{\lineheight{1.25}\smash{\begin{tabular}[t]{l}$(0,0,0)$\end{tabular}}}}%
    \put(0.12204234,0.82302548){\color[rgb]{0,0,0}\makebox(0,0)[lt]{\lineheight{1.25}\smash{\begin{tabular}[t]{l}$(0,0,1)$\end{tabular}}}}%
    \put(0,0){\includegraphics[width=\unitlength,page=1]{test-cube.pdf}}%
    \put(-0.09650992,0.88689296){\color[rgb]{0,0,0}\makebox(0,0)[lt]{\begin{minipage}{0.30117753\unitlength}\raggedright \end{minipage}}}%
  \end{picture}%
\endgroup%

%% file: figures/test-dodecahedron.pdf_tex
\begingroup%
  \makeatletter%
  \providecommand\color[2][]{%
    \errmessage{(Inkscape) Color is used for the text in Inkscape, but the package 'color.sty' is not loaded}%
    \renewcommand\color[2][]{}%
  }%
  \providecommand\transparent[1]{%
    \errmessage{(Inkscape) Transparency is used (non-zero) for the text in Inkscape, but the package 'transparent.sty' is not loaded}%
    \renewcommand\transparent[1]{}%
  }%
  \providecommand\rotatebox[2]{#2}%
  \newcommand*\fsize{\dimexpr\f@size pt\relax}%
  \newcommand*\lineheight[1]{\fontsize{\fsize}{#1\fsize}\selectfont}%
  \ifx\svgwidth\undefined%
    \setlength{\unitlength}{594bp}%
    \ifx\svgscale\undefined%
      \relax%
    \else%
      \setlength{\unitlength}{\unitlength * \real{\svgscale}}%
    \fi%
  \else%
    \setlength{\unitlength}{\svgwidth}%
  \fi%
  \global\let\svgwidth\undefined%
  \global\let\svgscale\undefined%
  \makeatother%
  \begin{picture}(1,0.94065657)%
    \lineheight{1}%
    \setlength\tabcolsep{0pt}%
    \put(-0.09650992,0.88689296){\color[rgb]{0,0,0}\makebox(0,0)[lt]{\begin{minipage}{0.30117753\unitlength}\raggedright \end{minipage}}}%
    \put(0,0){\includegraphics[width=\unitlength,page=1]{test-dodecahedron.pdf}}%
  \end{picture}%
\endgroup%

%% file: figures/test-endo-dodecahedron.pdf_tex
\begingroup%
  \makeatletter%
  \providecommand\color[2][]{%
    \errmessage{(Inkscape) Color is used for the text in Inkscape, but the package 'color.sty' is not loaded}%
    \renewcommand\color[2][]{}%
  }%
  \providecommand\transparent[1]{%
    \errmessage{(Inkscape) Transparency is used (non-zero) for the text in Inkscape, but the package 'transparent.sty' is not loaded}%
    \renewcommand\transparent[1]{}%
  }%
  \providecommand\rotatebox[2]{#2}%
  \newcommand*\fsize{\dimexpr\f@size pt\relax}%
  \newcommand*\lineheight[1]{\fontsize{\fsize}{#1\fsize}\selectfont}%
  \ifx\svgwidth\undefined%
    \setlength{\unitlength}{594bp}%
    \ifx\svgscale\undefined%
      \relax%
    \else%
      \setlength{\unitlength}{\unitlength * \real{\svgscale}}%
    \fi%
  \else%
    \setlength{\unitlength}{\svgwidth}%
  \fi%
  \global\let\svgwidth\undefined%
  \global\let\svgscale\undefined%
  \makeatother%
  \begin{picture}(1,0.94065657)%
    \lineheight{1}%
    \setlength\tabcolsep{0pt}%
    \put(-0.09650992,0.88689296){\color[rgb]{0,0,0}\makebox(0,0)[lt]{\begin{minipage}{0.30117753\unitlength}\raggedright \end{minipage}}}%
    \put(0,0){\includegraphics[width=\unitlength,page=1]{test-endo-dodecahedron.pdf}}%
  \end{picture}%
\endgroup%

%% file: figures/test-non-planar-dodecahedron.pdf_tex
\begingroup%
  \makeatletter%
  \providecommand\color[2][]{%
    \errmessage{(Inkscape) Color is used for the text in Inkscape, but the package 'color.sty' is not loaded}%
    \renewcommand\color[2][]{}%
  }%
  \providecommand\transparent[1]{%
    \errmessage{(Inkscape) Transparency is used (non-zero) for the text in Inkscape, but the package 'transparent.sty' is not loaded}%
    \renewcommand\transparent[1]{}%
  }%
  \providecommand\rotatebox[2]{#2}%
  \newcommand*\fsize{\dimexpr\f@size pt\relax}%
  \newcommand*\lineheight[1]{\fontsize{\fsize}{#1\fsize}\selectfont}%
  \ifx\svgwidth\undefined%
    \setlength{\unitlength}{594bp}%
    \ifx\svgscale\undefined%
      \relax%
    \else%
      \setlength{\unitlength}{\unitlength * \real{\svgscale}}%
    \fi%
  \else%
    \setlength{\unitlength}{\svgwidth}%
  \fi%
  \global\let\svgwidth\undefined%
  \global\let\svgscale\undefined%
  \makeatother%
  \begin{picture}(1,0.94065657)%
    \lineheight{1}%
    \setlength\tabcolsep{0pt}%
    \put(-0.09650992,0.88689296){\color[rgb]{0,0,0}\makebox(0,0)[lt]{\begin{minipage}{0.30117753\unitlength}\raggedright \end{minipage}}}%
    \put(0,0){\includegraphics[width=\unitlength,page=1]{test-non-planar-dodecahedron.pdf}}%
  \end{picture}%
\endgroup%

%% file: figures/architectures.tex
\begin{tabular}{lllll}
\toprule
Computing architecture& \\
\midrule
CPU \\         
  & vendor\_id	: AuthenticAMD \\
  & cpu family	: 23\\
  & model	: 24 \\
  & model name	: AMD Ryzen 7 PRO 3700U w/ Radeon Vega Mobile Gfx \\
& Output from "cpupower frequency-info": \\
&   "[CPU] Frequency should be within 2.30 GHz and 2.30 GHz. \\
&   The governor "performance" may decide which speed to use within this range." \\
Compiler \\
  & version : g++ 9.3.0-1 \\
  & optimization flags : -std=c++2a -O3 \\
\bottomrule
\end{tabular}

%% file: sections/conclusions.tex
\section{Conclusions}
\label{sec:concl}

\textcolor{Reviewer2}{A straightforward iterative algorithm is developed that significantly improves the computational efficiency of the VOF interface positioning problem on arbitrary unstructured meshes. The proposed Consecutive Cubic Spline (CCS) algorithm outperforms Brent's method, the stabilized secant-bisection method of \citet{Ahn2008}, and the Newton Cubic Spline method by \citet{Chen2019}. The CCS algorithm is comparable with the CIBRAVE method of \citet{Lopez2016,Lopez2018,Lopez2019} in terms of the average number of volume truncations, with a challenging volume fractions testing sequence and without relying on a relatively complex geometric parameterization of the truncated volume. Its relative simplicity and the usage of geometrical data that are already available in the geometrical VOF method simplifies the adoption of the proposed CCS algorithm in existing numerical codes.}

\section{Acknowledgments}

Funded by the Deutsche Forschungsgemeinschaft (DFG, German Research Foundation) – Project-ID 265191195 – SFB 1194, sub-project Z-INF. The author is very grateful to Jan-Patrick Lehr M.Sc. for his advice on improving the performance measurements.